\documentclass[manuscript,screen]{acmart}
\usepackage{makecell}
\usepackage{graphicx}
\usepackage{subfigure}
\usepackage{float}
\usepackage{multirow}
\usepackage{graphicx}
\usepackage{gensymb}
\usepackage{multirow,tabularx}
\usepackage{threeparttable}
\usepackage{makecell}
\usepackage{array}
\usepackage{nomencl}

\newcolumntype{P}[1]{>{\centering\arraybackslash}p{#1}}
\newcolumntype{M}[1]{>{\centering\arraybackslash}m{#1}}
\AtBeginDocument{%
  }

\setcopyright{acmlicensed}
\copyrightyear{2025}
\acmYear{2025}
\acmDOI{XXXXXXX.XXXXXXX}

\acmISBN{978-1-4503-XXXX-X/18/06}

\makenomenclature

\begin{document}

\title{Immersive Multimedia Communication:\\
State-of-the-Art on eXtended Reality Streaming}


\author{Haopeng Wang}
\affiliation{%
 \institution{University of Ottawa}
 \city{Ottawa}
 \state{Ontario}
 \country{Canada}}
 \email{hwang266@uottawa.ca}

\author{Haiwei Dong}
\affiliation{%
  \institution{University of Ottawa and Huawei Canada}
  \city{Ottawa}
  \state{Ontario}
  \country{Canada}}
\email{hdong@uottawa.ca}

\author{Abdulmotaleb El Saddik}
\affiliation{%
  \institution{University of Ottawa}
  \city{Ottawa}
  \state{Ontario}
  \country{Canada}}
\email{elsaddik@uottawa.ca}

\renewcommand{\shortauthors}{Wang et al.}

\begin{abstract}
Extended reality (XR) is rapidly advancing, and poised to revolutionize content creation and consumption. In XR, users integrate various sensory inputs to form a cohesive perception of the virtual environment. This survey reviews the state-of-the-art in XR streaming, focusing on multiple paradigms. To begin, we define XR and introduce various XR headsets along with their multimodal interaction methods to provide a foundational understanding. We then analyze XR traffic characteristics to highlight the unique data transmission requirements. We also explore factors that influence the quality of experience in XR systems, aiming to identify key elements for enhancing user satisfaction. Following this, we present visual attention-based optimization methods for XR streaming to improve efficiency and performance. Finally, we examine current applications and highlight challenges to provide insights into ongoing and future developments of XR.
\end{abstract}



\keywords{Do, Not, Us, This, Code, Put, the, Correct, Terms, for,
  Your, survey}

\received{20 February 2007}
\received[revised]{12 March 2009}
\received[accepted]{5 June 2009}

\maketitle

\section{Introduction}
The term Extended Reality (XR) \nomenclature{XR}{Extended Reality} refers to a broad category of technologies that are aiming to merge physical and virtual worlds to provide immersive, interactive experiences to the user, which includes virtual reality (VR) \nomenclature{VR}{Virtual Reality}, augmented reality (AR) \nomenclature{AR}{Augmented Reality}, and mixed reality (MR) \nomenclature{MR}{Mixed Reality}\cite{9754241}. In recent years, XR has fundamentally transformed our life in various aspects, including work, education, social interaction, and entertainment, by seamlessly integrating the physical and digital realms. The XR market is forecasted to reach 1,706.96 billion USD by 2032, with a compound annual growth rate of 32.1\% from 2024 to 2032 \cite{market}. The rapid advancement of hardware and software technologies has accelerated the expansion of the XR market, greatly improving the accessibility and effectiveness of immersive experiences \cite{8493267}. For instance, XR experiences are becoming more accessible to a wider range of individuals thanks to the increasing number of smartphones and wearable devices that feature XR features. Moreover, due to the worldwide transition to remote work and digital communication caused by the COVID-19 pandemic, there has been a significant increase in the need for remote collaboration tools. As a result, XR solutions are becoming increasingly popular in both consumer and enterprise sectors. 

The XR offers fascinating user experiences by allowing unrestricted movement and seamless interaction with both physical and virtual environments in real-time. Since XR applications aim to deliver immersive experiences, the perceived QoE (quality of experience) \nomenclature{QoE}{Quality of Experience} is paramount for XR users. In order to improve existing services and develop future services, the QoE is a crucial metric to evaluate and understand users' expectations and experiences \cite{ruan2021survey}. However, the assessment of QoE metrics for XR systems remains a significant challenge because there are a number of factors from different disciplines that contribute to QoE. Furthermore, guaranteeing a good QoE in XR systems requires a significant amount of storage space, computation capability, and network bandwidth. One of the most essential issues is the exponential growth of content traffic, which imposes severe challenges on current network infrastructures. The widespread adoption of XR applications has further escalated the demands for superior network quality and performance. Additionally, XR technology introduces new challenges in system design, dynamic viewpoint prediction and adaptive streaming.

Previous studies often regard XR as a subset of broader multimedia technologies, focusing primarily on applications in specific domains such as education \cite{su132413776}, healthcare \cite{andrews2019extended}, industry \cite{cardenas2022extended}, and engineering \cite{ANASTASIOU2023100105}. However, these works typically offer limited attention to the distinct challenges associated with XR streaming. Additionally, several surveys have investigated 360° video streaming \cite{CHEN201866,CHIARIOTTI2021133,8960364,zink2019scalable,8756213}. For instance, Chen et al. \cite{CHEN201866} provide a comprehensive review of omnidirectional video coding, focusing on projection techniques and their impact on video quality. Xu et al. \cite{8960364} review developments in 360° video and image processing, emphasizing visual attention modeling, quality assessment, and compression techniques. Zink et al. \cite{zink2019scalable} analyze 360° video streaming systems, addressing content creation, storage, distribution, rendering, QoE evaluation, and edge-based distribution models. Given XR’s objective to deliver high Quality of Experience (QoE), several studies focus specifically on image and video quality assessment \cite{min2024perceptual,ruan2021survey,10384639}. For example, Duan et al. \cite{10384639} review visual and multimodal attention modeling and perceptual quality assessments in XR environments. Min et al. \cite{min2024perceptual} survey quality assessment approaches across streaming, VR/AR, and user-generated content. Ruan et al. \cite{ruan2021survey} investigate QoE evaluation methods for VR streaming, emphasizing machine learning-based QoE optimization techniques. However, these studies predominantly address the challenges of 360° video streaming from a QoE perspective, leaving broader XR streaming challenges relatively underexplored.

This paper provides a comprehensive survey of current advancements, challenges, and methodologies associated with XR streaming. It identifies gaps in prior research, which focused mainly on 360-degree video or specific applications, with limited attention to XR streaming challenges. By emphasizing the unique requirements of XR streaming and the need for specialized research, this paper provides an in-depth examination of multimodal interactions, traffic patterns, and adaptive streaming technologies. We provide the definitions of XR terms, including AR, VR, and MR, followed by a typical XR streaming system architecture and a detailed analysis of XR traffic features in Section \ref{2}. Section \ref{3} summarizes the multimodal interaction techniques used in popular XR devices. Section \ref{4} discusses key factors influencing QoE. In Section \ref{5}, we introduce the primary visual-attention optimization approaches at both the application and network layers. Key applications and challenges are discussed in Sections \ref{6} and \ref{7}. Finally, we summarize the survey in Section \ref{8}.

\section{Overview of Extended Reality (XR) Systems and Traffic} \label{2}
\subsection{Definitions of XR Technologies}
\begin{figure}[t]
  \centering
  \includegraphics[width=\linewidth]{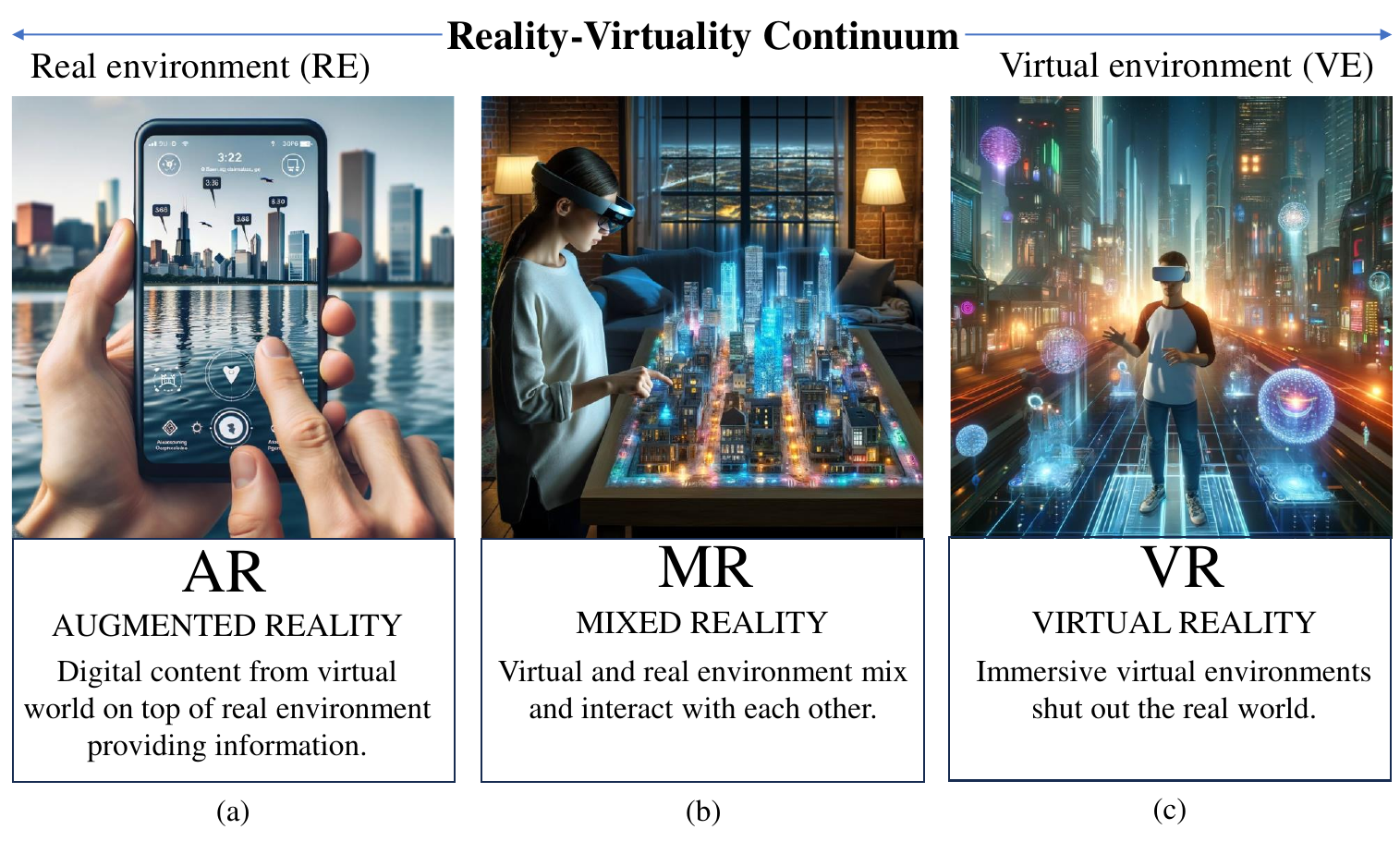}
  \caption{Definition of XR Technologies: AR, MR, and VR according to the Reality-Virtuality Continuum (the figures are generated by LLM). (a) A user interacts with a city using AR on a mobile phone. (b) A user wearing a HoloLens 2 MR headset interacts with a city displayed on a table.(c) A VR user stands in a city and interacts with it in a fully immersive way.}
  \label{XR}
\end{figure}

XR covers a variety of immersive environments that combine the physical and digital worlds using advanced computing and human-machine interaction. The "X" in XR could stand for different spatial computing technologies \cite{10.1145/2756547}. 
Even though XR has the potential to integrate more technologies, our survey mainly concentrates on VR, AR, and MR. As shown in Fig. \ref{XR}, the definitions and relationships of XR technologies can be explained using the reality-virtuality continuum \cite{milgram1995augmented}, which outlines a spectrum spanning from an exclusively physical reality to an entirely virtual worlds, offering users varied levels of immersion and interactivity. The introductions of VR, AR, and MR are outlined in detail below.

\begin{itemize}
    \item \textit{Virtual Reality (VR):} VR technologies cover the virtual environment at the end of the reality-virtuality continuum. VR enables the creation of a fully immersive digital environment, where the real-world surroundings are entirely obscured. By wearing a VR headset or head-mounted display (HMD) \nomenclature{HMD}{Head-Mounted Display}, users are able to see a 360-degree view of an artificial world. This immersive experience creates a convincing illusion that tricks the brain into perceiving that users are in a new and dynamic environment. It enables users to explore and engage with virtual environments and objects in a highly realistic and captivating way.
    
    \item \textit{Augmented Reality (AR):} AR technologies cover an area close to the real environment in the reality-virtuality continuum. AR enables the overlay of digital elements onto the physical world. This encounter enhances the physical world by integrating digital elements such as images, text, and animations. Users can access these elements via AR glasses, tablets, and smartphones. While there may be a certain level of interaction between physical and virtual elements in specific AR experiences, direct interaction between digital and physical components is usually restricted or entirely absent. 
    
    \item \textit{Mixed Reality (MR):} MR technologies occupy the center of the reality-virtuality continuum. It superimposes digital features onto the real world, allowing physical and digital items to coexist and interact with each other in real-time. Consequently, MR systems receive input from the environment and adapt accordingly. For instance, users can place digital objects within the room they are in, rotate them, or interact with these virtual elements in various ways, creating an engaging and interactive experience.

\end{itemize}

\begin{figure}[htbp]
  \centering
  \includegraphics[width=\linewidth]{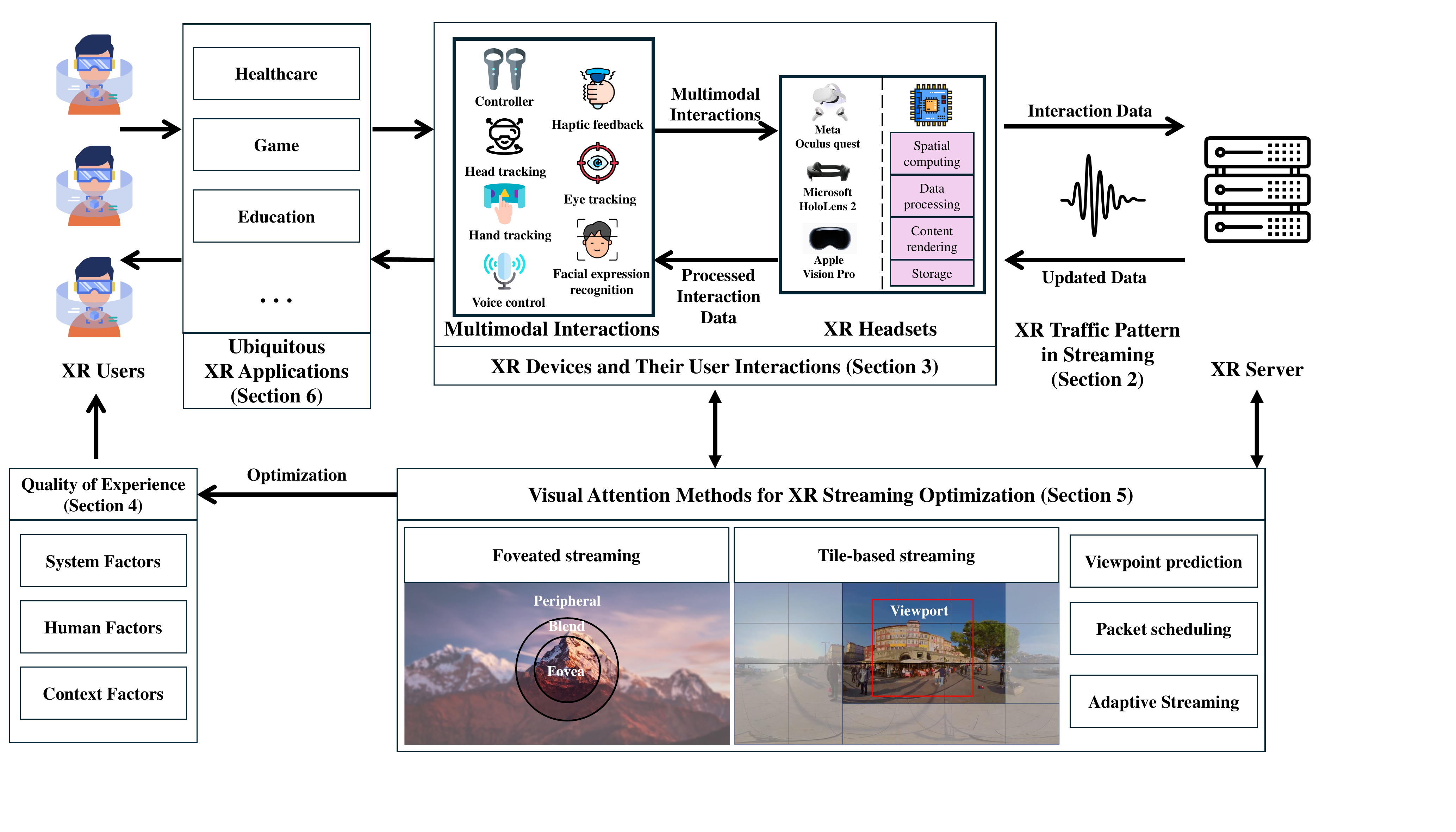}
  \caption{The architecture of an XR system comprises users, XR devices, a streaming network, and an XR server. Users engage with XR headsets through various multimodal interaction techniques, while QoE optimization involves methods applied to the application and network layers.}
  \label{xr system}
\end{figure}

\subsection{XR System Architecture}
% \section{XR System Architecture}
A typical schematic diagram of an XR system is presented in Fig.\ref{xr system}, which contains XR users, headsets, and a server. As most current XR systems adopt local rendering, the XR headsets are responsible for most computing and processing workloads, such as user input capturing, content rendering, spatial computing, and data processing (e.g., local application logic, and algorithms). The local application logic refers to a set of rules and operations executed on a local device. 
The XR server performs complex calculations for XR application mechanics, manages the XR application's global logic, processes inputs from all connected headsets, and resolves any conflicts to maintain a consistent application state and ensure that all users are experiencing the same content. By distributing real-time updates to clients, the server ensures that all users have a synchronized view of the virtual world, allowing for cohesive and engaging multiplayer experiences.

The users interact with the XR system through various multimodal interaction methods (described in section \ref{3}) via input devices and sensors. The acquired interaction data is processed by XR headsets or other devices and sent to the XR server. The XR server provides content updates to the XR headsets. The headsets render content and process interaction data, transmitting the results back to the users.
The communication between the server and headsets occurs in real time, ensuring that actions by one user are promptly and accurately reflected in the XR environment for others. The XR headsets and server together enable a seamless and immersive environment where users can engage in complex, real-time interactions within a consistent virtual world.

\subsection{XR Traffic Pattern in Streaming}
Before discussing the optimization methods for XR streaming, we first explore the XR traffic pattern to discover the potential issues in XR streaming. Since XR offers users multimodal interaction and immersive experiences, its traffic differs significantly from traditional content traffic. Despite growing interest and notable progress in XR, the characterization of traffic in XR streams is yet mainly unclear. There has been relatively little work on analyzing and modeling XR network traffic. Thus, XR systems require advancements in evaluating traffic on current communication systems to guarantee state-of-the-art performance and QoE for users.

\begin{figure}[htbp]
  \centering
  \includegraphics[width=\linewidth]{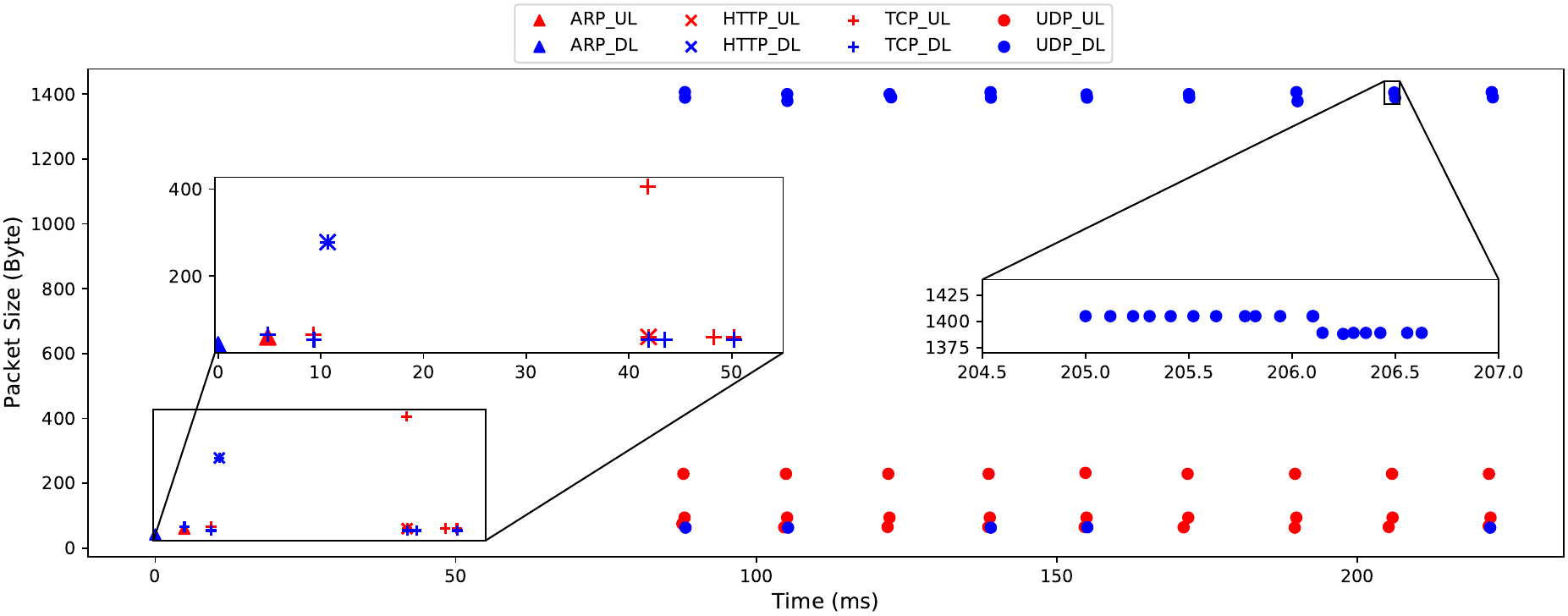}
  \caption{An example of a traffic pattern for an XR platform. The stream is divided into two stages: the connection stage and the transmission stage. During the connection stage, HTTP is used, while UDP is employed during the transmission stage.}
  \label{traffic}
\end{figure}

As shown in Fig. \ref{traffic}, XR traffic can be split into connection and transmission stages. HTTP is employed during the connection stage to ensure a stable connection, while UDP is used during the transmission stage to minimize delay. Meanwhile, the traffic can be divided into downlink (DL) \nomenclature{DL}{Downlink} and uplink (UL) \nomenclature{UL}{Uplink} streams. Various types of packets are present in both streams. More precisely, a typical HTTP session occurs during the connection stage. In the transmission stage, the UL stream includes packets for synchronization, interaction data and video frame reception information, while the DL stream comprises video frame packet bursts, synchronization, and acknowledgment. The primary data component in the XR DL stream is the video frame, which is transmitted in large packet bursts. The UL interaction information is the second most significant traffic stream. This information is collected by the XR devices and then sent to the server to refresh the content. Furthermore, smaller packets have been detected in both the UL and DL streams. These packets serve as feedback regarding the reception of video frames. This feedback is likely utilized in the streaming protocol to determine the necessity of retransmitting frames \cite{lecci2021open}.

Baldoni et al. \cite{10.1145/3625468.3652187} introduce a dataset named Questset, which was collected using the Quest 2 headset from 70 participants over more than 40 hours. Through evaluation of this dataset, the authors discovered that video frames are streamed in packet bursts, with the frame rate serving as the inter-frame interval. They also observed instances of skipped frames, where certain frames could be predicted from preceding frames, thus obviating the need for re-rendering or transmission. This phenomenon results in a larger inter-frame interval. 

Traffic patterns of various social VR platforms are also evaluated \cite{10302989, 10.1145/3517745.3561417, 10430223}. After loading the VR model and establishing the connection, the XR system streams data over UDP, showing strong periodicity and regularity in packet distribution. Since the VR content is loaded upon access and rarely changes, streamed packets primarily handle connection, synchronization, acknowledgment, and interaction. The data transfers exhibit burstiness, with the amount of exchanged data being relatively small. Social VR platforms support numerous interactions that typically require low bandwidth. However, the bandwidth requirement increases significantly when new models are loaded, leading to delays and extended waiting times. In addition, the download throughput experiences a linear increase as new users join the platform, which may lead to scalability concerns. Meanwhile, the DL includes both content traffic and real-time multimedia signals, such as voice and video. Real-time performance is degraded when the capacity of the access network is exceeded by the DL rate.

The aforementioned research indicates that existing XR systems provide lower-quality experiences with relatively minimal computational and bandwidth requirements due to limitations in rendering and network infrastructure \cite{10.1145/3625468.3652187, 10302989, 10.1145/3517745.3561417, 10430223}. However, as XR technology develops, there is an anticipated shift towards applications designed to provide high-fidelity, fully immersive experiences. These advanced XR applications place significantly higher demands on network quality and computational capabilities to meet elevated user expectations \cite{10173785, 10299766, 10007756, 9984845}. While current XR systems may operate with moderate computational power and bandwidth, advancements in HMDs, such as improved resolution, field of view, and refresh rates, can significantly enhance the realism and responsiveness of XR environments. These hardware improvements will enable more seamless and visually compelling immersive experiences, though they will also heighten the demand for higher bandwidth to handle high-resolution, low-latency content. Therefore, while upgrading HMDs is crucial, research focused on optimizing network infrastructure and bandwidth is essential to support next-generation XR applications, enabling real-time interactions without compromising quality. For instance, achieving high-fidelity immersion requires the ideal end-to-end XR system delay to be less than 7 ms, corresponding to the duration of the vestibulo-ocular reflex process, i.e., 7 ms \cite{9984845}. For 360-degree XR content, the required bandwidth could reach 2.3 Tbps, considering a 360 $\times$ 180 degree field of view, 64 pixels per degree (PPD) \nomenclature{PPD}{Pixels Per Degree}, a frame rate of 30 FPS and 8-bit color depth \cite{10007756}.

\section{XR Devices and Their User Interactions} \label{3}
Many tech giants have invested significant efforts in XR technologies and released many commercial products such as the Apple Vision Pro, Microsoft HoloLens 2, Meta Oculus Quest 3, Google Glass, Samsung Gear VR, and HTC Vive. Although various devices, including smartphones, computers, tablets, and headsets, support XR applications, headsets are the most dominant and immersive devices. The technical features of various XR headsets are listed in Table \ref{xr headsets}. We also provide detailed introductions to three popular XR headsets: Apple Vision Pro, Microsoft HoloLens 2, and Meta Oculus Quest 3. 

\subsection{State-of-the-Art XR Headsets}
\subsubsection{Apple Vision Pro}
The Apple Vision Pro, released on June 5, 2023, is a mixed-reality headset developed by Apple \cite{apple}. This device utilizes physical inputs for interaction, including hand tracking, eye tracking, voice recognition and facial expression recognition, and operates on VisionOS, which is built upon iOS frameworks.
The headset is equipped with dual 4K micro-OLED displays, presenting a total of 23 million pixels and typically operates at 90 FPS (frames per second) \nomenclature{FPS}{Frames Per Second}. Moreover, it can automatically adjust to 96 or 100 FPS depending on the content displayed. The Apple Vision Pro includes an extensive array of cameras and sensors: six world-facing tracking cameras, four eye-tracking cameras, two high-resolution main cameras, a TrueDepth camera for facial recognition, and a LiDAR Scanner for depth mapping. Additionally, the device is fitted with a flicker sensor, four inertial measurement units (IMUs), and an ambient light sensor to enhance user interaction and environmental integration.
The Vision Pro employs two processors to power these advanced features. The Apple M2 chip, known for its powerful graphics capabilities, supports the VisionOS and handles complex computer vision algorithms. In contrast, the newly introduced Apple R1 chip processes inputs from the device's cameras, sensors, and microphones, ensuring rapid image transmission within just 12 milliseconds. This dual-processor setup allows the Vision Pro to deliver sophisticated 3D experiences in a mixed-reality context.

\nomenclature{IMU}{Inertial Measurement Unit}
\nomenclature{LLMs}{Large Language Models}
\subsubsection{Microsoft HoloLens 2}

The HoloLens 2, developed by Microsoft, is an advanced mixed-reality headset released on November 7, 2019 \cite{hololens}. This device enhances user interaction through a variety of intuitive inputs including hand tracking, eye tracking, and voice recognition. It runs on the Windows Holographic operating system, which is based on Windows 10. Equipped with a see-through holographic display, the HoloLens 2 uses a 2k 3:2 light engine to deliver a more immersive visual experience with a greater field of view (FoV) \nomenclature{FoV}{Field of View} than its predecessor. The system is designed to render holograms within the user's physical environment, offering a blend of the virtual and real worlds. The device includes several sensors and cameras to support a wide range of functionalities: a 1-megapixel time-of-flight depth sensor, an 8-megapixel camera for capturing both images and videos, and an array of MR capture cameras. It also features an accelerometer, gyroscope, and magnetometer, which are crucial for spatial recognition and navigation within the mixed-reality environment. The HoloLens 2 is powered by the Qualcomm Snapdragon 850 Compute Platform, which handles both the processing of holographic data and the overall operation of the Windows Holographic OS. This integration allows for efficient handling of complex computations and real-time data processing, facilitating a seamless interactive experience.

\subsubsection{Meta Oculus Quest 3}
The Meta Quest 3, developed by Meta (formerly Facebook), is an all-in-one VR headset that was released in 2023 \cite{Meta}. The Quest 3 advances the frontiers of immersive VR experiences through significant enhancements in both hardware and software. The device is equipped with a dual-LCD display, delivering a combined resolution of 2064 x 2208 pixels per eye, thereby offering superior visual sharpness and an expanded field of view, which collectively enhance the perceived depth and clarity of virtual environments. Powered by the Qualcomm Snapdragon XR2 Gen 2 Platform, the Quest 3 boasts improved processing power and efficiency, enabling higher frame rates and more detailed graphics in VR applications. The headset operates predominantly at 120 FPS, providing smooth and responsive visual performance, with adaptive refresh rates that adjust according to the content. Furthermore, the Quest 3 introduces advanced hand tracking and enhanced haptic feedback in its Touch Controllers, contributing to a more tactile and interactive user experience. Similar to its predecessor, the Quest 3 employs inside-out tracking via multiple integrated cameras, facilitating seamless movement within a user's physical space without the necessity for external sensors. The device also utilizes the updated Quest Platform, which is based on an enhanced version of Android, offering a more refined user interface and an expanded content library. These advancements in display technology, processing capability, and user interaction underscore the Meta Quest 3's role as a significant advancement in the effort to make high-fidelity VR experiences more accessible to a broader audience.

\begin{table}[htbp]
\centering
\caption{Comprehensive Specifications and Interaction Ways for Popular XR Devices.}
\begin{threeparttable}
\begin{tabular}{|p{3cm}|p{0.8cm}|p{1.5cm}|p{1cm}|p{1.2cm}|p{5cm}|}
\hline
\textbf{Device}                   & \textbf{Type} & \textbf{Resolution} & \textbf{FoV}\tnote{1}  & \textbf{FPS \tnote{2}} & \textbf{Interaction Ways}                                                     \\ \hline
Oculus Quest 3                   & VR            & 2064 x 2208                 & 118°                           & 120      & Hand tracking, controllers, voice recognition, head tracking                         \\ \hline
HTC Vive Pro 2                    & VR            & 2448 x 2448                 & 113.30°                         & 120     & Hand tracking, controllers, eye tracking, head tracking, peripherals             \\ \hline
HTC Vive XR Elite                    & MR            & 1920 x 1920                 & 110°                         & 90     & Hand tracking, controllers, eye tracking, head tracking, voice commands, gesture control             \\ \hline
Valve Index                       & VR            & 1440 x 1600                 & 114.43°                         & 144  & Hand tracking, controllers, head tracking, peripherals                           \\ \hline
Microsoft HoloLens 2              & MR            & 2048 × 1080               & 52°                           & 60           & Hand tracking, eye tracking, voice commands, gesture control, head tracking       \\ \hline
Magic Leap 1                      & AR            & 1280 x 960                  & 50°                           & 120          & Hand tracking, eye tracking, voice commands, gesture control, head tracking       \\ \hline
Sony PlayStation VR 1              & VR            & 960x1080                  & 100°                         & 120      & Controllers, head tracking, voice recognition, peripherals                           \\ \hline
Pico Neo 3                        & VR            & 1832 x 1920                 & 113.08°                           & 90       & Hand tracking, controllers, voice recognition, head tracking                         \\ \hline
Apple Vision Pro                  & MR            & 3660x3142               & 100°                            & 100       & Eye tracking, hand gestures, voice recognition, head tracking, facial expression tracking \\ \hline
Varjo XR-3                        & MR         & 2880 x 2720                 & 106°                          & 90           & Eye tracking, hand tracking,  haptic feedback, head tracking                  \\ \hline
Samsung Gear VR                   & VR            & Depends on the smartphone   & 110°                                & Depends      & Controllers, head tracking, depends on smartphone sensors,                          \\ \hline
Google Glass Enterprise Edition 2 & AR            & 640 x 360                   & 83°                          & --           & Voice recognition, head tracking                                                      \\ \hline
Epson Moverio BT-40             & AR            & 1920x1080                  & 34°                           & 60           & Head tracking, controllers, voice recognition                                                      \\ \hline
Nreal Air                       & AR            & 1920x1080                       & 46°                           & 60           & Controlled by smartphone                                                \\ \hline
\end{tabular}
\begin{tablenotes}
    \footnotesize               %这行要添加
        \item[1] FoV represents field of view.  
        \item[2] FPS represents frames per second.  
\end{tablenotes}
\end{threeparttable}
\label{xr headsets}
\end{table}

The XR device market offers a diverse range of products tailored to meet various needs. High-end devices, such as the Apple Vision Pro and Microsoft HoloLens 2, provide premium MR experiences with state-of-the-art processing power and advanced tracking technologies for hand, eye, and facial tracking, making them ideal for enterprise applications. Mid-range devices, such as the Oculus Quest 3 and HTC Vive Pro 2, feature high-resolution displays and wide fields of view, optimized for gaming and media consumption, though they lack the full MR capabilities of their high-end counterparts. Entry-level devices, including the Magic Leap 1 and Google Glass Enterprise Edition 2, focus on lightweight designs and moderate resolution, targeting industrial applications such as augmented overlays and remote assistance. The evolution of XR technology is characterized by advancements in display resolution, wider FoV, increased processing power, and sophisticated multimodal interaction features, such as eye tracking, hand tracking, voice recognition, and haptic feedback. Furthermore, the transition to standalone, wireless devices makes XR technology more immersive, accessible, and versatile, paving the way for broader adoption and integration into everyday life.

\subsection{Multimodal Interactions}
XR headsets typically support multimodal interaction technologies beyond visual and audio, enabling users to experience virtual environments through hand tracking, voice commands, gaze tracking, and haptic feedback. While many experimental and emerging technologies hold potential for interaction, they are not yet ready for deployment in XR applications. In this section, we introduce key multimodal interaction technologies, excluding basic visual and audio interactions, as follows:

\begin{itemize}
    \item \textit{Controller}: The controller is an essential component of an XR system that enables users to engage with the virtual world. An XR controller is equipped with buttons, thumbsticks, triggers, and sensors designed to monitor hand movements and convert them into virtual actions. The device functions as a tool that extends the capabilities of the hand by enabling users to control things, navigate through menus, and execute various activities within the virtual environment \cite{YUAN202337}.
    
    \item \textit{Head Tracking}: Head tracking is a fundamental mechanism for interacting with XR systems. It involves tracking the user's head movements and orientation to provide a more immersive and responsive experience using a combination of sensors and algorithms. By accurately capturing head movements, XR systems can adjust the visual and auditory output to match the user's perspective, thereby enhancing the sense of presence and immersion in the virtual environment. 
    
    \item \textit{Hand Tracking}: Hand tracking and gesture recognition enable users to engage with virtual worlds by utilizing their hands, avoiding the need for traditional controllers. The XR systems utilize cameras and sensors to track hand or finger movements to interpret gestures such as pinching, grabbing, and swiping. For example, users can pick up, move, or resize items in a virtual space with this method \cite{buckingham2021hand}.
    
    \item \textit{Voice Recognition}: Voice recognition technology enables users to control XR systems and interact with virtual elements using voice commands. Built-in microphones capture the user's voice, and the speech-processing algorithm interprets the commands.
    
    \item \textit{Haptic Feedback}: Haptic devices provide physical feedback through vibrations or forces to simulate the touch or interaction with virtual objects, enhancing the realism of virtual experiences. Many XR controllers and haptic devices, such as gloves or suits, are equipped with advanced feedback mechanisms including vibration motors and force feedback systems, all designed to greatly enhance the immersive experience.
    
    \item \textit{Eye Tracking}: Eye tracking monitors the user's gaze direction and allows interactions based on where they are looking. Eye movements are tracked by sensors and cameras, and algorithms are used to trigger actions based on the data. Eye tracking can be used to control interfaces, enhance immersion, and optimize rendering techniques. 
    
    \item \textit{Facial Expression Recognition}: Facial expression recognition is a technology that allows the system to detect and interpret the user's facial expressions. This interaction method enhances the immersive experience by enabling avatars or digital characters to reflect the user's emotions in real time, adding a layer of realism and personal connection to virtual interactions. As technology advances, this capability will become increasingly integral to applications across social XR, gaming, training, and mental health, creating more engaging and effective virtual experiences. 
\end{itemize}

\section{Quality of Experience} \label{4}
As XR aims to provide users with an immersive experience, the perceived QoE is crucial. For the development of current and future XR services, it is essential to understand the user experiences and expectations. Many foundational metrics for evaluating QoE in XR originate from traditional video assessments, focusing on resolution, bitrate, frame rate, buffering, and color depth \cite{app12157581, ruan2021survey}. While these metrics establish a baseline for XR QoE, their application in XR is more complex and dynamic. Developing a comprehensive QoE model remains a significant challenge due to XR's interactive and immersive nature. As shown in Table \ref{factors}, the factors influencing QoE can be categorized into three groups \cite{ 10.1007/978-3-031-15101-9_12, ruan2021survey, hal-04638470}: system factors, which encompass intrinsic system properties affecting the experience, such as hardware capabilities, network QoS (quality of service) \nomenclature{QoS}{Quality of Service} parameters, and media configurations; context factors, which include external environments in which the system operates, such as physical location, social context, and specific use case scenarios; and human factors, which relate to physiological and psychological perceptions that humans have of the experience, including sensory input, cognitive load, and emotional responses.

\begin{table}[htbp]
\centering
\caption{The Factors Influencing QoE.}
\label{factors}
\begin{tabular}{|c|c|M{3cm}|M{7cm}|}
\hline
Factors              & Type & Sub-factors                 & Examples \\ \hline

\multirow{4}{*}{System factors} & \multirow{4}{*}{Objective} & Network factors                             & Latency, throughput, packet loss, buffering event rate, buffering time, bitrate, bandwidth \cite{7965610, 10056951, 10.1145/2018436.2018478}         \\ \cline{3-4} 
                               & & Application factors                     &   Resolution, frame rate \cite{ghinea2005quality}       \\ \cline{3-4} 
                               & & Service factors                          &  Content type \cite{duan2017assessment}, viewing mode, application level    \cite{8509121}    \\ \cline{3-4} 
                               & & Hardware factors                         &  HMDs, smartphones, interaction, FoV, frame rate  \cite{7965658,8509121}      \\ \hline
\multirow{2}{*}{Human factors} & \multirow{2}{*}{Subjective}  & Physiological factors            & Gender,  vision, hearing \cite{6178834, saleme2021influence}        \\ \cline{3-4} 
                               & & Psychological factors              &  Background, education level, preference, mood \cite{wechsung2011all, silvia2008interest, kortum2010effect}        \\ \hline
                               
\multirow{2}{*}{Context factors}  & \multirow{2}{*}{Objective}      & Physical factors                    &  Lighting \cite{9946999},  sound, location \cite{han2012qoe, 10632616}       \\ \cline{3-4} 
                               & & Economic factors                    &   Desired price, budget \cite{sackl2017more, sackl2014got, sackl2014evaluating}       \\ \hline
\end{tabular}%

\end{table}

\subsection{System Factors} 
A system factor is a quality or characteristic that impacts the overall performance of an XR service or application in terms of technical criteria. System factors are grouped into four categories: network, application, service, and hardware factors. Network factors refer to network characteristics influencing the delivery and performance of XR content to users, such as packet loss, system latency, throughput, average bit rate, buffering time, buffering event rate, and network bandwidth. Application factors encompass technical specifications and settings defining how content is processed and presented, such as resolution and frame rate. Service factors are attributes associated with the content and user interaction that influence user engagement and enjoyment, such as the type of content being viewed, the complexity of the application, and the selected viewing mode. Hardware factors pertain to the physical components and capabilities of XR devices, such as HMDs, headphones, decoder performance, head-tracking technology, and FoV. Each of these categories significantly impacts the overall quality of experience for the user. Many of these factors, such as resolution, frame rate, and latency, originate in traditional video QoE research, where are primarily evaluated in passive viewing scenarios. In XR, however, they are expanded to address additional complexities, such as real-time interaction and low-latency requirements, to ensure synchronization between the virtual experience and user movements.

The content quality has a straightforward impact on the QoE. A user's experience can also be adversely affected by other issues stemming from algorithms and hardware, such as blockiness and blur. Additionally, system factors such as media configurations \cite{1468162} and network QoS parameters \cite{10.1145/2018436.2018478} significantly impact QoE. Dobrian et al. \cite{10.1145/2018436.2018478} conduct a study in which they assess various quality metrics, including buffering ratio, rendering quality, join time, average bitrate, and rate of buffering events. They discover that the buffering ratio, representing the proportion of time spent in buffering, is the most important factor influencing user engagement across all content types.
Ghinea et al. \cite{ghinea2005quality} check the impact of color depth and frame rate, and find that users' satisfaction and understanding of the presentation are not proportionally diminished by significant frame loss or color depth reduction. Zhang et al. \cite{8509121} propose a QoE evaluation framework including four high-level parameters: hardware quality, content quality, user interaction, and environment understanding.
Singla et al. \cite{7965658} evaluate the impact of HMD devices and user behaviors on the QoE. 
The research undertaken by \cite{8463413} investigates the trajectory and speed of both head movement and object movement in VR. In addition, this study examines several aspects of the content, such as the complexity of the background, to acquire a deeper understanding of how users perceive it under different circumstances. They also collect assessments of sickness levels from a total of 80 participants. Regarding the perceptual quality of 360-degree video, Shahid et al. \cite{8710766} conduct a subjective evaluation of the effects of content type, encoding parameters, and rendering device on QoE while considering the user's profile. Their findings indicate that viewers exhibit greater tolerance towards encoding parameters when watching engaging 360-degree videos in VR, compared to less engaging content. Additionally, the study reveals that device type significantly impacts viewer satisfaction, with higher mean opinion scores recorded for content viewed on HTC Vive compared to Google Cardboard.

With the development of deep learning, neural network models are designed to evaluate the factors influencing QoE. Duan et al. \cite{10056951} introduce a deep learning-based metric to detect critical distortion that impacts VR image quality, such as color mismatches, blurring, and ghosting. 
Zhu et al. \cite{9961939} propose an approach to assess in-the-wild image quality without a reference, capturing both semantic and distortion-specific details. Liu et al.\cite{yang2024aigcoiqa2024} evaluate AI-generated omnidirectional images based on quality, comfortability, and correspondence. The quality measures visual fidelity including sharpness and color, while the comfortability evaluates the user’s immersive experience by assessing image realism and structural coherence. The correspondence checks the alignment between the generated image and its guiding text prompt. Zhu et al. \cite{zhu2024esiqa} use a subjective quality assessment method where human subjects rate the perceptual quality of egocentric spatial images. Sun et al. \cite{8702664} propose a multi-channel CNN model for no-reference quality assessment of 360-degree images. Duan et al. \cite{8351786} employ a subjective quality evaluation method that gathers human ratings on the perceptual quality of omnidirectional images viewed in a VR environment. They further 
 investigate the influence of factors such as visual oscillations, immersion duration, and video content \cite{duan2017assessment}. Moreover, they evaluate the impact of different parameters, such as resolution, bit rate, and frame rate, on video quality in VR environments \cite{7965610}.

\subsection{Human Factors}
Human factors in XR QoE build on insights from traditional video QoE research. Metrics for assessing user comfort, satisfaction, and perceptual responses, are applicable in XR but require expansion to account for the heightened sensory and cognitive demands. Both human physiological and psychological elements have a substantial impact on QoE \cite{ruan2021survey}.

\subsubsection{Physiological Factors}
The physiological factors, such as gender, age, and other physiological characteristics, play a crucial role in QoE. Laghari et al. \cite{6178834} analyze various factors inherent to the human body, such as gender and age, to identify the primary influences on user perception quality. While many of these elements have been extensively investigated and modeled, the specific impact of an individual’s physiological characteristics on QoE remains a vital area of investigation. Saleme et al. \cite{saleme2021influence} study 360° mulsemedia (multiple sensorial media), an emerging XR application, to investigate the physiological aspects that could influence the experience. In contrast to previous research, the authors introduce odor sensitivity as a distinct variable and discover that women had greater sensitivity in scenarios involving several senses. 
Shahid et al. \cite{8710766} also investigate QoE in XR through the analysis of user profile data, including age, gender, interest in the content and familiarity with panoramic VR content, alongside other parameters such as encoding settings, content type, and device type. Their findings indicate that users have a higher level of tolerance towards encoding rates when viewing engaging 360-degree panoramic VR videos and are less sensitive to encoding rates than when viewing less engaging content. Additionally, viewers showed a marked preference for certain device types. Consequently, by analyzing user profiles, content service providers and device manufacturers can efficiently allocate resources to deliver services that meet user expectations.
 
\subsubsection{Psychological Factors}
It has been demonstrated that the psychological state of the user significantly affects QoE in various ways \cite{wechsung2011all, moller2014quality, kortum2010effect, palhais2012quality}. The study by Palhais et al. \cite{palhais2012quality} shows that viewers tend to overlook video quality issues when they are interested in the content, indicating a positive correlation between interest and QoE. Additionally, other psychological factors such as personality, attitudes, motivation, attention levels, and mood also play crucial roles in influencing QoE \cite{wechsung2011all}. Some studies identify interest as a key influencing factor in QoE \cite{silvia2008interest, kortum2010effect}. This interest can be triggered by specific content, thereby significantly affecting the user's perception of QoE.

\subsection{Context Factors}
Contextual factors encompass the situational characteristics that define a user's surroundings. While contextual factors are applicable in traditional video QoE studies, they gain heightened relevance in XR, where users interact within more immersive and variable environments. Factors such as lighting, sound, and location, traditionally assessed in passive contexts, become critical in XR due to their direct influence on immersion and user comfort. These factors can differ in their magnitude, behaviors, and patterns of occurrence, both on their own and as a group. These elements are categorized into physical environmental aspects (e.g., lighting, sound, and location) and economic factors (e.g., pricing preferences and budget restrictions).

Han et al. \cite{han2012qoe} argue that the QoE of a user is affected by multiple external elements present in their surrounding environment. They find that when users are relaxed, their QoE improves. Additionally, the user's experience is substantially influenced by physical factors such as the location of the seat, the distance and height of the viewing area, the lighting conditions, and potential disruptions such as incoming calls or notifications from short message services \cite{staelens2010assessing}. Martinez et al. emphasize economic contextual factors, such as the cost of subscription, as influential in QoE. Yamori et al. \cite{yamori2004relation} find that the amount a user pays for content affects their experience, with users generally exhibiting higher tolerance for content with lower prices. Furthermore, studies conducted by Sackl et al. \cite{sackl2017more, sackl2014got, sackl2014evaluating} reveal that incorporating factors, such as financial constraints, user expectations, and pricing based on quality, contribute to the performance of user perception models. Duan et al. \cite{9946999} evaluate the impact of real-world contextual factors, such as lighting and background complexity, on the perceptual quality of superimposed AR imagery. They later propose a framework for evaluating image quality in AR environments using visual confusion theory, which examines the effects of superimposing digital content on real-world scenes on \cite{9828671}.
Wang et al. \cite{10632616} examine how the interplay between virtual and real worlds affects perceptual quality.

\section{Visual Attention Methods for XR Streaming Optimization} \label{5}

\nomenclature{MAE}{Mean Absolute Error}

\subsection{Visual Attention}
In XR systems, users typically view scenes within a limited FoV and focus on the most attractive and interesting areas. Despite the wide FoV of the human visual system, the highest visual acuity is concentrated in the foveal region, which spans only the central 2.5° of the visual field \cite{jabbireddy2022foveated}. Leveraging this feature, numerous optimization methods have been proposed to reduce bandwidth usage and computational power. The core idea behind these methods is to identify where a user is viewing or which parts are more visually attractive and likely to be viewed. Consequently, the XR system streams content near the user's viewpoint in high quality while delivering other areas in lower quality. As these methods utilize features of the human visual system, we refer to these techniques as visual attention methods and categorize them as foveated streaming and tile-based streaming. Foveated streaming divides the screen into foveal, blend, and peripheral regions \cite{soler2017proposal}. The foveal region, where visual acuity is highest, is streamed at the highest resolution, aligning with the user’s gaze. The blend region serves as a transitional area with medium resolution, ensuring smooth detail transitions. The peripheral region, with the lowest visual acuity, is streamed at a reduced resolution, leveraging the eye’s insensitivity to detail in this area to conserve computational resources. This technique is particularly suited for real-time applications such as gaming, virtual simulations, and XR workspaces, as these scenarios involve rapid gaze shifts and dynamic interactions. The adaptive nature of foveated streaming ensures high visual quality in focus areas while minimizing latency and computational demands, both critical for maintaining responsiveness in these environments. Tile-based streaming divides XR content into rectangular tiles. High-quality streams are delivered for tiles within the visible viewport, while tiles outside the viewport are streamed at lower quality. This method prioritizes bandwidth for regions actively observed by the user, ensuring efficient resource utilization and enhancing the viewing experience \cite{shafi2020360}. It is especially effective for 360° video streaming, remote collaboration, and virtual tourism, where user viewing patterns are more predictable, and content is often pre-rendered. These characteristics allow tile-based streaming to optimize resource utilization, enhance immersion, and deliver consistent quality without the need for real-time adaptation.

\nomenclature{KLD}{Kullback-Leibler Divergence}
\nomenclature{PCC}{Pearson Correlation Coefficient}
\nomenclature{NSS}{Normalized Scanpath Saliency}

\subsection{Viewpoint Prediction}
Both foveated rendering and tile-based streaming utilize gaze to determine the area where a user is looking. Despite the capabilities of real-time eye-tracking, There is a natural delay between when a specific point of gaze is detected by the eye tracker and when the corresponding visual content is updated in the HMD frame. This latency can negate any quality improvements achieved through optimization methods such as foveated rendering and tile-based streaming. Moreover, it can adversely impact the QoE by optimizing regions that are no longer under foveal vision \cite{10.1145/3534086.3534331}. As a result, methods for predicting future gaze positions have gained prominence. 

\begin{figure}[htbp]
    \centering
    \subfigure[Viewport Prediction: Historical (green), predicted (red), and ground truth (blue) viewport scanpaths for viewport prediction.]{
        \includegraphics[width=0.7\textwidth]{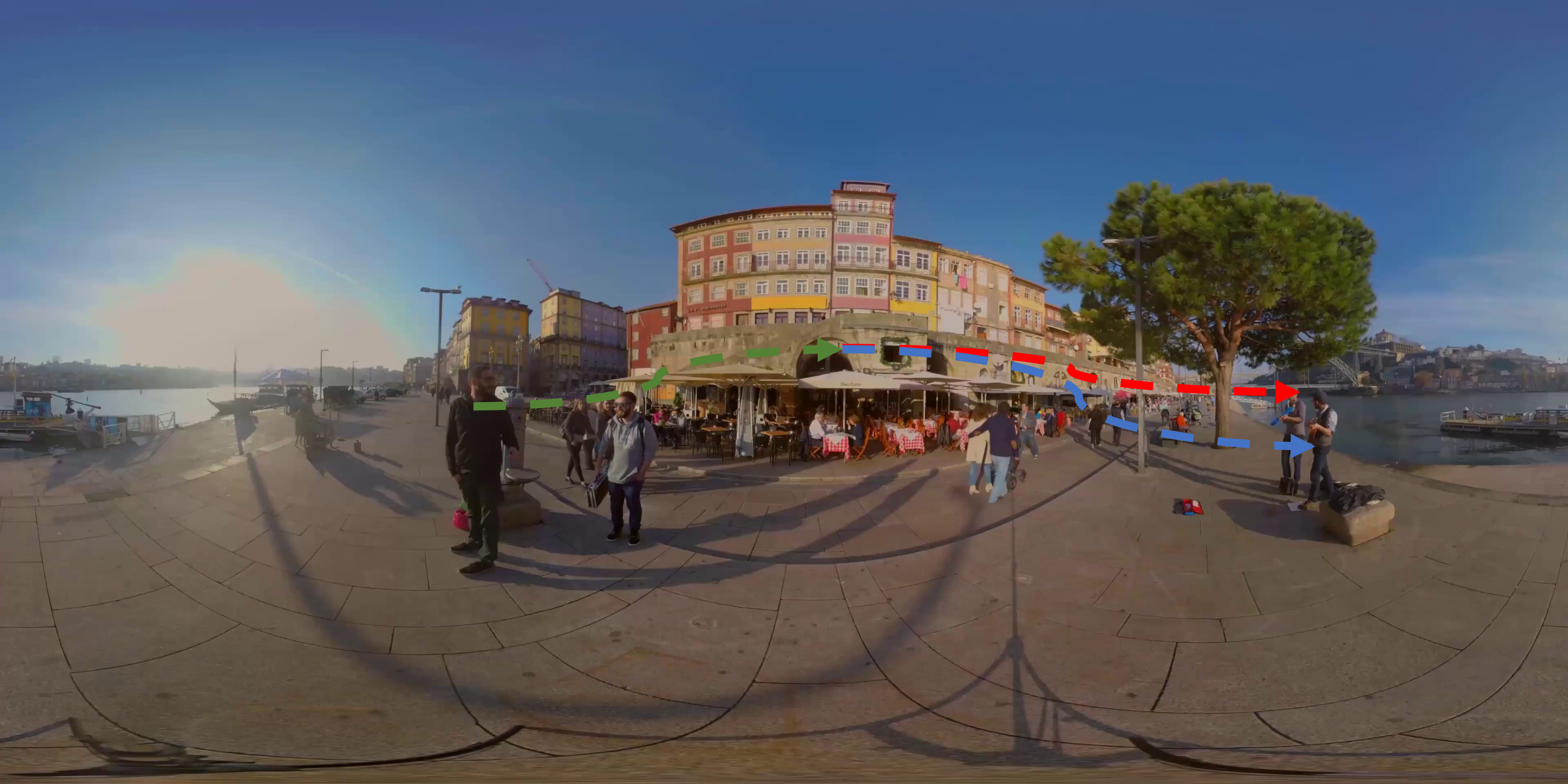}
        \label{viewport_prediction}
    }
    \hspace{1cm} % Adjust the spacing
    \subfigure[Saliency prediction: Heatmaps showing attention regions predicted in 360° content.]{
        \includegraphics[width=0.7\textwidth]{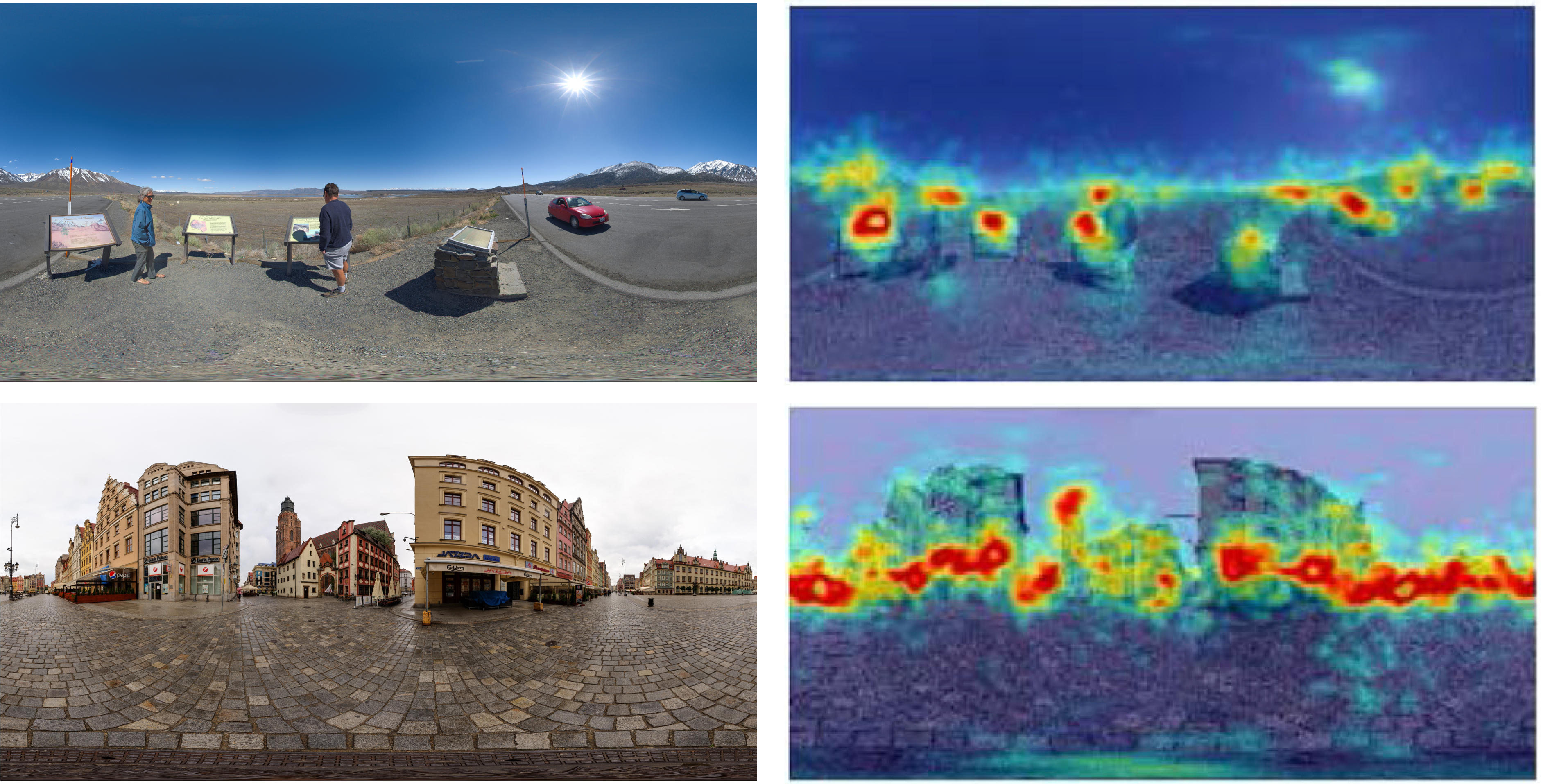}
        \label{saliency_map}
    }
    \caption{Illustrations of viewpoint prediction methods.}
    \label{viewpoint_prediction}
\end{figure}

As shown in Fig. \ref{viewpoint_prediction}, viewport prediction forecasts the specific region of a video or scene a user is likely to observe in the near future, typically centered on the anticipated gaze direction. Saliency prediction identifies regions most likely to draw visual attention due to perceptual or cognitive significance, irrespective of the user’s current gaze focus. Viewport prediction methods are evaluated using metrics, such as mean absolute error (MAE), great circle distance, and overlap accuracy are commonly applied for viewport prediction \cite{9733647,9234071}. MAE calculates the average deviation between predicted and actual viewpoints, while great circle distance measures the shortest path between predicted and actual viewpoints on a spherical surface. Overlap accuracy measures the fraction of the actual viewport area covered by the predicted viewport. For saliency prediction, widely used metrics include Kullback-Leibler divergence (KLD), Pearson correlation coefficient (PCC), and normalized scanpath saliency (NSS) are widely used \cite{8931644,10.1145/3503161.3547955, 10.1145/3083187.3083218}. KLD measures the distribution difference between predicted and true saliency maps, where lower values reflect closer alignment. PCC assesses the linear correlation between predicted and actual saliency distributions. NSS compares predicted saliency with human fixation points, indicating alignment with user attention. These metrics collectively evaluate viewpoint and saliency prediction methods, balancing directional precision, spatial overlap, and alignment with actual user focus to optimize XR content delivery. Although predicting where a user will look is often referred to as viewport prediction in the literature, a more precise term is viewpoint prediction. Viewpoint refers to the center of the viewport, which is determined by the yaw and pitch angles. Similarly, the goal of saliency prediction aim to predict the position that the human eye pays attention to in images or videos. In this survey, we summarize saliency and viewport prediction methods, using the terms interchangeably as they serve the shared goal of optimizing content delivery based on anticipated user focus. Over the last decades, numerous viewpoint and saliency prediction methods have been proposed, broadly categorized into classical machine learning-based and deep learning-based approaches.

% \usepackage{multirow}
% \usepackage{graphicx}
\begin{table}[]
\centering
\caption{Summary of Viewpoint Prediction Methods for XR Systems.}
\label{vp methods}

\begin{tabular}{|M{2.5cm}|M{1.5cm}|M{3cm}|M{4cm}|M{1.5cm}|}
\hline
               Methods                              & Citation                                       & Input type         & Algorithms                                                                 & Prediction Horizon \\ \hline
\multirow{5}{*}{\begin{tabular}[c]{@{}l@{}}Classical Machine \\ Learning Methods\end{tabular}}         & Qian et al. \cite{qian2016optimizing}      & Historical trajectory    & Weighted linear regression                                                 & 4s                 \\ \cline{2-5} 
                                             & Hu et al. \cite{hu2019vas360}          & Historical trajectory                & Weighted linear regression                                                 & 2s                 \\ \cline{2-5} 
                                             & Ban et al. \cite{ban2018cub360}           & Historical trajectory                & LR and KNN                                                                     & 6s                  \\ \cline{2-5} 
                                             & Petrangeli et al. \cite{8613652}                 & Historical trajectory                & clustering algorithm and trend trajectory function                                                    & 10s                \\ \cline{2-5} 
                                             & Xie et al. \cite{10.1145/3123266.3123291} & Historical trajectory         &   clustering algorithm under Gaussian distribution assumption                                          & 3s                  \\ \hline
\multirow{9}{*}{\begin{tabular}[c]{@{}l@{}}Deep Learning\\  Methods\end{tabular}} & Xu et al. \cite{8578657}                 & Saliency map and video frames     & CNN and LSTM                                                                  & 1s                \\ \cline{2-5} 
                                             & Hu et al. \cite{8998375}                 & Historical trajectory and video frames      & CNN                                                                        & 1s                \\ \cline{2-5} 
                                             & Fu et al. \cite{9234071}                 &  Historical trajectory                & LSTM model with cross-attention mechanism                                                    & 5s                 \\ \cline{2-5} 
                                             & Chao et al. \cite{9733647}                 &  Historical trajectory                & Transformer encoder                                                        & 5s                 \\ \cline{2-5} 
                                             & Xu et al. \cite{8418756}                 & Historical trajectory and video frames & DRL                                                                        & 30ms               \\ \cline{2-5} 
                                             & Nguyen et al. \cite{10.1145/3240508.3240669} & Historical trajectory and saliency maps    & CNN and LSTM                                                                 & 2.5s               \\ \cline{2-5} 
                                             & Rondón et al. \cite{9395242}                 & Historical trajectory and video frames     & Stack LSTM model                                                                      & 5s                 \\ \cline{2-5} 
                                             & Guimard et al. \cite{10.1145/3524273.3528176} & Historical trajectory               & Discrete variational multiple sequence model based on deep latent variable model & 5s                 \\ \cline{2-5} 
                                             & Wang et al. \cite{10525060}                & Historical trajectory and video frames      & Multimodal temporal-spatial transformer                                                                & 5s                 \\ \cline{2-5}
                                              &  Zhu et al.\cite{9606880} & Historical trajectory and video frames            & Graph-Based CNN & --                 \\ \cline{2-5} 
                                            &  Duan et al. \cite{10.1145/3503161.3547955} & AR image, background image, and superimposed image            & Vector Quantized Encoder-Decoder model & --                 \\ \cline{2-5} 
                                            &  Zhu et al. \cite{zhu2024does} & Video frames and audio clip          & U-Net architecture & 2s                 \\ \hline 
\end{tabular}%

\end{table}

\subsubsection{Classical Machine Learning-based Methods}
Various linear regression (LR) \nomenclature{LR}{Linear Regression} algorithms are used by several existing methodologies to predict future viewing positions with historical viewpoint trajectories \cite{qian2016optimizing, hu2019vas360}. Additionally, some probabilistic models have been proposed to estimate the distribution of prediction errors to enhance the performance of linear regression methods \cite{10.1145/3123266.3123291,8351404}. However, LR-based methods assume linear head movement, which is a strong assumption that introduces significant bias. Consequently, numerous methods have been developed to extract spatial and temporal features from different users' viewpoint trajectories, achieving better performance and becoming predominant in existing XR streaming systems. Since viewpoint trajectories of an application from various users exhibit similar spatial and temporal characteristics, the user's viewpoint trajectory can be predicted based on historical data from other users with clustering methods \cite{ban2018cub360, liu2017360}. A spectral clustering approach is used to categorize trajectories that share similarities \cite{8613652}. For each cluster, a specific function is computed to predict future viewpoint positions. Similarly, Taghavi et al. \cite{10.1145/3386290.3396934} clustered viewpoint trajectories from previous users into different groups. By extrapolating the quaternions, the user's trajectory is matched to one of these clusters, and viewpoints are predicted using the cluster center.

\subsubsection{Deep Learning-based Methods} 
Many deep learning-based methods have been proposed to predict user viewpoints. Hu et al. \cite{8998375} focus on dynamic scenes, predicting future gaze positions with a CNN (convolutional neural network) -based model. Meanwhile, LSTM (long short-term memory) \nomenclature{LSTM}{Long Short-Term Memory} networks are widely employed for viewpoint prediction \cite{10.1145/3534086.3534331, 9024132, 9234071}. Xu et al. \cite{8578657} build a dataset of gaze data from observers in dynamic 360-degree content and use CNN and LSTM networks for gaze displacement prediction. For instance, Fu et al. \cite{9234071} combine LSTM with a self-attention mechanism to predict viewpoints. Zhang et al. \cite{9024132} construct three LSTM models and use the mean of their predictions as the final result.  As the transformer \cite{10.5555/3295222.3295349} has made progress in many fields, it is also used in viewpoint prediction. Chao et al. \cite{9733647} utilize the transformer encoder to predict viewpoints.
To enhance the accuracy of viewpoint prediction, additional information, such as saliency maps and video content, is integrated into deep learning models. Xu et al. \cite{8418756} present a deep reinforcement learning (DRL) method to predict head movements. This method takes 360-degree video content and past viewport trajectories as input and optimizes the difference between the agent's actions and the user's movements. Romero et al. \cite{9395242} develop an LSTM model that leverages past viewpoint trajectories and saliency maps to forecast future viewpoints. Nguyen et al. \cite{10.1145/3240508.3240669} propose a CNN architecture to predict saliency maps and an LSTM model to predict future viewpoints using these predicted saliency maps and head orientation maps. 
Zhu et al. \cite{10.1145/3565024} introduce a visual behavior adaptive saliency model to enhance saliency prediction by integrating spatial-temporal cues and visual behavior adaptations using a Markov chain-based algorithm. Later, they present two methods: a graph-based viewing behavior model and a graph-based CNN model, both utilizing head and eye movement data to enhance saliency prediction by addressing projection distortions and leveraging spatial-temporal information \cite{9606880}. Furthermore, they propose a saliency prediction model for 360-degree images, which uses spherical harmonics to capture features across different frequency bands, combining low-level visual features and high-level cues to generate accurate saliency maps for head and eye movements \cite{8931644}. Duan et al. \cite{10.1145/3503161.3547955} develop a vector quantized saliency prediction model tailored for AR scenes, based on eye-tracking data. Zhu et al. \cite{10.1007/978-3-031-46317-4_29} demonstrate the significant influence of audio, particularly ambisonic sound, on user focus in VR environments. In addition, they propose an audio-visual saliency prediction network that hierarchically fuses audio and visual features within a multimodal aligned embedding space \cite{zhu2024does}.

\nomenclature{DRL}{Deep Reinforcement Learning}
\nomenclature{CNN}{Convolutional Neural Network}

All the aforementioned methods focus on predicting a single-viewpoint trajectory. However, Guimard et al. \cite{10.1145/3524273.3528176} highlight the necessity for multiple-viewpoint prediction by analyzing public viewpoint data, given the various possible future trajectories that can arise from similar past trajectories. To address this, they propose a discrete variational learning method for multiple-viewpoint predictions. Similarly, Wang et al. \cite{10525060} develop a transformer-based method to predict multiple-viewpoint trajectories along with their viewing probabilities by treating viewpoint prediction as a classification problem. These approaches aim to capture the inherent uncertainty and variability in user behavior, providing a more comprehensive and accurate prediction model for future viewpoints.

The viewpoint prediction techniques for both traditional and deep learning-based methods are summarized in Table \ref{vp methods}. It is evident that the most essential data for viewpoint prediction is historical trajectory. As deep learning advances, more types of information are being used to improve viewpoint prediction accuracy, including saliency maps and video frames.

\subsection{Adaptive Streaming}
The aforementioned foveated rendering and viewport streaming techniques optimize XR content spatially by dividing the content into smaller areas, with the areas closest to the viewpoint streamed in high quality while others are streamed in lower quality. Additionally, the content quality can be dynamically adjusted temporally to further reduce bandwidth and computational requirements. Adaptive streaming is a crucial application-layer technology for optimizing content delivery in XR applications by dynamically adjusting media quality according to network conditions and device capabilities. 
This ensures a smooth and immersive user experience by minimizing buffering and playback interruptions. By encoding XR content at multiple quality levels and adjusting in real-time, adaptive streaming efficiently uses bandwidth, enhances scalability, and improves accessibility. 

\begin{figure}[h]
  \centering
  \includegraphics[width=\linewidth]{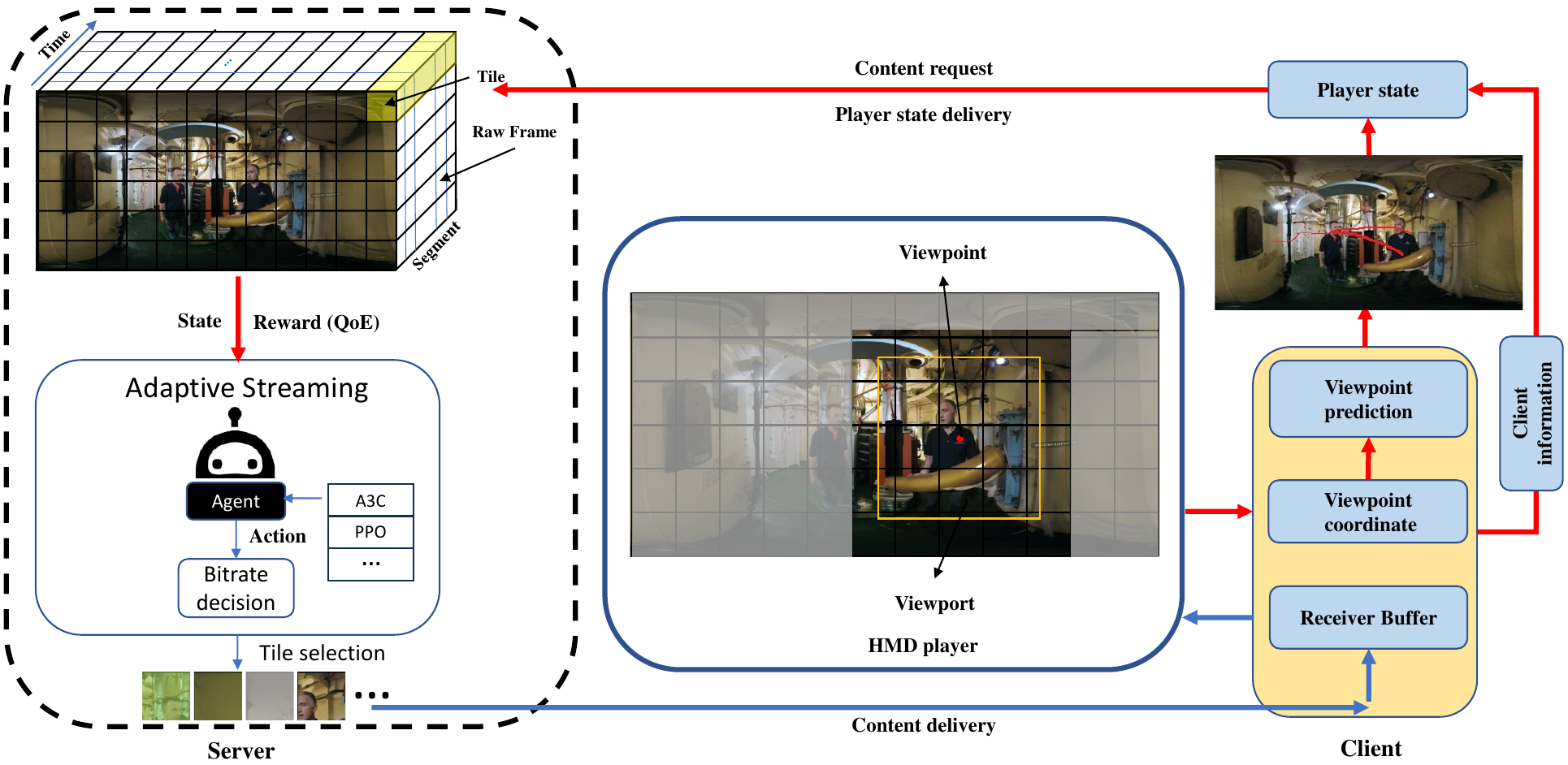}
  \caption{A typical framework of visual-attention-based adaptive streaming method. The XR content is split into many areas. The quality of each area is determined by various DRL algorithms, such as A3C, PPO.}
  \label{adaptive streaming}
\end{figure}

Adaptive streaming is typically formulated as a QoE optimization problem, which is an NP-hard problem \cite{9234071}. Consequently, various heuristic algorithms have been proposed, such as beam search \cite{9024132}, dynamic programming \cite{10.1145/3240508.3240556}, and greedy algorithms \cite{10.1145/3123266.3123291}. Hu et al. \cite{9301338} formulate XR streaming as a convex optimization problem to maximize the user's QoE and used the dichotomy method to obtain the optimal solution. Zhang et al. \cite{8576614} formulated QoE maximization as an NP-hard problem and proposed a ranking-based heuristic solution to determine each tile's quality based on its priority. Nevertheless, these heuristic solutions are time-consuming and face difficulties in achieving optimal outcomes under diverse network circumstances. 

Various methods based on single-agent deep reinforcement learning (SADRL) \nomenclature{SADRL}{Single-Agent Deep Reinforcement Learning} have been proposed to address these challenges. Given the increasing dimension of the action space with the number of tiles and bitrate levels, Fu et al. \cite{9234071} present an adaptive streaming strategy that sequentially determines bitrate for each tile with the Asynchronous Advantage Actor-Critic (A3C) \nomenclature{A3C}{Asynchronous Advantage Actor-Critic} algorithm \cite{DBLP:journals/corr/MnihBMGLHSK16}. The complexity of the action space can be further reduced by adjusting the bitrate based on the viewport region. Zhang et al. \cite{8737361} divided the VR video into two regions: viewport and rest. The tiles within the viewport are allocated the same bitrate, which is determined by a SADRL model with A3C. Conversely, the remaining tiles are assigned the lowest bitrate. Tang et al. \cite{9035396} streamed the entire VR video without using a tile-based method to multiple users and adopted the SADRL method with the A3C algorithm to make bitrate decisions for each user to maximize QoE. Kan et al. \cite{9419061} split the VR video into three regions: viewport, marginal, and invisible, and presented a SADRL model with the A3C algorithm to determine the bitrates for these three regions simultaneously. Wei et al. \cite{9351629} proposed a two-step strategy to determine the bitrate of tiles. A SADRL model first determines the segment bitrate, and then the bitrates of individual tiles are determined using game theory, considering view prediction and segment bitrate. Feng et al. \cite{9838819} classified the tiles inside the viewport into different levels and utilized the Proximal Policy Optimization (PPO) \nomenclature{PPO}{Proximal Policy Optimization} algorithm to determine the bitrate for these tiles. Long et al.\cite{10144631} propose an adaptive resource allocation approach to assign communication and computation resources based on multi-agent deep reinforcement learning (MADRL) and graph convolutional networks for multiple users.

However, all existing methods based on SADRL usually achieve local optima for bitrate determination without globally considering the presence of other tiles. Therefore, Wang et al. \cite{10525060} formulate XR streaming as a decentralized partially observable Markov decision process (Dec-POMDP) optimization problem and propose a MADRL \nomenclature{MADRL}{Multi-Agent Deep Reinforcement Learning} method using the multi-agent proximal policy optimization (MAPPO) \nomenclature{MAPPO}{Multi-Agent Proximal Policy Optimization} algorithm to globally determine the bitrate for tiles based on multi-viewpoint predictions from the transformer method.

As shown in Fig. \ref{adaptive streaming}, adaptive streaming can be formulated as a QoE optimization problem and solved by DRL. The XR content is split into many tiles, and the quality of each tile is determined by the DRL algorithm by observing the environment state. \nomenclature{RL}{Reinforcement Learning} The state, action and reward of the DRL model are shown below:

\begin{itemize}
    \item \textit{State}: The RL agent takes a state of the environment after playing n frames, including but not limited to predicted viewpoint position, download time for the past n frames, network throughput for the past n frames, the quality of the last frame, and the current buffer level.
    \item \textit{Action}: The action of the RL agent is bitrate. As bitrate is a continuous action (scalar), many DRL methods can be applied, such as PPO, A3C, etc. 
    \item \textit{Reward}: A designed XR QoE model can be used as the reward for the RL. For example, a QoE model could consist of four main components:
\begin{equation}
    Q_t = Q_t^1 - \eta_1 \cdot Q_t^2 - \eta_2 \cdot Q_t^3 - \eta_3 \cdot Q_t^4
\end{equation}
where $Q_t^1$ is viewport quality at time step $t$ which represents the average quality of video content within the user's viewport.
$Q_t^2$ is viewport temporal Variation, which measures the change in quality between consecutive viewports.
$Q_t^3$ is viewport spatial variation, which accounts for rate changes among tiles within the user's viewport to prevent blocking artifacts.
$Q_t^4$ is rebuffering time representing the impact of buffering events on user experience.
The $\eta_*$ are adjustable parameters that allow for different user preferences. This model aims to balance high viewport quality against minimizing variations and rebuffering events.
\end{itemize}

\subsection{Packet Scheduling}
The streaming of XR over a time-varying network is a complex problem involving many variables and parameters. Several network-adaptive packet scheduling algorithms exist for traditional video streaming, ranging from basic methods, such as tail drop and priority scheduling, to more complex mechanisms designed to achieve fairness and minimize tail latency \cite{10.1145/2834050.2834085}. The tail drop algorithm is a simple and popular packet scheduling algorithm, which is widely used in practice. Packets are dropped from the tail of the queue when traffic congestion occurs. However, the tail drop does not differentiate between packet types. 

Hence, priority scheduling algorithms are proposed for traditional video streaming. One popular priority packet scheduling method is to drop packets based on frame type, such as I-frame (intra-frame), P-frame (predictive frame), and B-frame (bi-directional frame). In \cite{ Cha2003}, video frames are dropped randomly based on the priority labels applied to the I, P, and B frames. Gobatto et al. \cite{https://doi.org/10.48550/arxiv.2202.04703} propose a packet drop algorithm to avoid IRAP (intra random access point)-packet loss. If network congestion is detected, non-IRAP packets could be preemptively dropped until the congestion disappears. These research works, however, do not take into account differences between frames of the same type. Typically, the first P frame in a group of pictures (GOP) \nomenclature{GOP}{Group of Pictures} causes more distortion than subsequent P frames. As a result, frames of the same type have varying effects on the quality of reconstruction.

\nomenclature{I-frame}{Intra-Frame}
\nomenclature{P-frame}{Predictive Frame}
\nomenclature{B-frame}{Bi-Directional Frame}
\nomenclature{IRAP}{Intra Random Access Point}

Therefore, more sophisticated methods are proposed for modeling the impact of packets on video quality and generating packet scheduling schemes. 
Chakareski et al. \cite{1608103} propose an optimization framework to solve the problem of packet scheduling for multiple videos over a limited network link. Video packets are characterized using rate-distortion information. By discarding packets, a distributed streaming technique enables a trade-off between rate and distortion among many streams. However, their work aims to achieve fairness between multiple traditional videos instead of VR videos. Corbillon et al. \cite{7574700} prioritize traditional video packets using an evaluation function with taking into account frame type, frame dependency and frame size instead of the quality distortion. Meanwhile, packets are filtered in the order according to their importance obtained from the evaluation function.

\begin{table*}[htbp]
\caption{Comparison of the existing packet scheduling methods}
\centering
\label{one}
\begin{threeparttable}
\begin{tabular}{|M{2cm}|M{1.3cm}|M{1.3cm}|M{0.5cm}|p{4cm}|p{3.7cm}|}
\hline
Citation    & Content Type  & Resolution   & FPS\tnote{1}& \makecell[c]{Formulated Problem}    & \makecell[c]{Proposed Method}   \\ \hline
Chakareski et al. \cite{1608103} & Traditional Content & 176 × 144    & 30  & Utilize rate-distortion information to maximize the overall quality of multiple videos streaming over a limited bandwidth transmission channel & Compute Lagrange multiplier of nonconstrained optimization problem using gradient method with Lagrangian relaxation  \\ \hline
Corbillon et al. \cite{7574700}        & Traditional Content & 1920 × 1080   & 25  & Optimize the degradation of video by considering the type, size, and dependency of frames  & Using the evaluation function, drop frames based on their importance    \\ \hline
Nasralla et al. \cite{NASRALLA2018126} & Traditional Content & 640 × 416     & 25  & Reduce packet delay and improve quality using a utility function based on frame type and temporal complexity,   & Discard packets based on their priority as determined by the utility function          \\ \hline
Change et al. \cite{6172571}           & Traditional Content & 720 × 480     & 30  &  Reduce the visual impact of frame loss by minimizing the visual score & Drop B-frame based on visual score  \\ \hline
Cosma et al. \cite{9068423}            & XR Content          & ---          & --- & Improve the fraction of time in the transmission time interval by allocating the available frequency resources to different traffic classes  & Reinforcement learning with continuous actor-critic learning automata algorithm    \\ \hline
Wei et al.    \cite{9500817}              & XR Content          & 3840 × 2048   & 30  & Reduce the distortion of the viewport over multiple paths by selecting bitrate based on bandwidth and delay of each path                       & The water filling algorithm gradually allocates flow from the path with the least delay   \\ \hline
Chakareski \cite{9086630}              & XR Content          & 3840 × 2048                                                            & 30  & Maximum VR video quality delivered from multiple base stations  considering content popularity, rate-distortion, and base station information & A faster iterative algorithm is used to obtain an approximate solution to the problem                                                                                  \\ \hline
Ge et al. \cite{7997740} & XR content & --- & --- & Minimize a network latency model containing five latency factors &  A software-defined networking architecture and a multi-path cooperative route scheme are posed to reduce network latency \\ \hline
Wang et al. \cite{10494038}                                                    & XR Content          & 3840 × 1920                                                             & 25  & Minimize the distortion of the entire XR Content and viewport over limited bandwidth transmission channel considering viewport and rate-distortion         & The optimization problem is solved using dynamic programming containing state transition equations and initial states                                                                       \\ \hline
\end{tabular}%
\begin{tablenotes}
    \footnotesize               %这行要添加 
        \item[1] FPS represents frames per second.  
\end{tablenotes}
\end{threeparttable}
\end{table*}

% Chou et al. \cite{1608118} present a flexible framework for controlling packet transmissions in a network with packet loss based on rate-distortion optimized methods. The issue of rate-distortion optimized streaming can be simplified by analyzing the transmission of an isolated packet with an emphasis on minimizing error costs. This involves taking into account the deadlines, interdependencies, and distortions that may occur if the packets are not received. A gradient descent algorithm is proposed to solve the optimization problem. However, the framework is proposed for traditional video instead of VR video. 
Nasralla et al. \cite{NASRALLA2018126} propose a content-aware packet scheduling method for video streaming. A suggested utility function prioritizes packets for video transmission depending on the temporal complexity and kind of frames, such as I frames, P frames, and B frames. In the system, packets are dropped based on their prioritization. However, their work ignores the interdependencies of frames and the rate information. Kang et al. \cite{Kang2002} shows a packet scheduling algorithm where video packets of different importance are scheduled by using different deadline thresholds. A packet's importance depends on its motion-texture context and relative position in its GOP. Video packets are scheduled based on the deadline threshold differently from the order in which they were originally played. 

% Chang et al. \cite{6172571} investigate the visual impact of whole frame loss and find that the system still achieves high quality with frame dropping to reduce the bitrate. They developed a visibility model for B-frame loss to generate a visual score for each frame before video transmission. Frames are dropped based on visual scores to reduce data volume. However, only B-frame is allowed to be dropped in their system. 

In spite of the rapid development of XR, very few packet scheduling methods have been proposed for XR content transmission in recent years. Cosma et al. \cite{9068423} introduce a packet scheduling technique that utilizes machine learning to distribute network resources for real-time VR video and other media applications. RL is used to prioritize different traffic classes and determine a packet scheduling rule. However, their work solves the problem of resource allocation between multiple traffics from VR video and other applications instead of data reduction for a VR video. Meanwhile, it is possible that a single path may not meet the demanding specifications of VR videos. VR video streaming on multipath simultaneously is proposed to improve VR video quality. Wei et al. \cite{9500817} propose a VR video streaming framework based on multipath TCP. The framework dynamically selects the bitrate for the viewport according to the network conditions of all paths, such as delay, and packet loss. The system schedules video packets in different paths to deliver VR video on time. However, their system is designed to select the bitrate in accordance with network conditions and to allocate resources across multiple paths. Chakareski \cite{9086630} presents a joint transmission system based on multiple cell base stations for VR video. Based on historical viewport data, a statistical model is proposed for determining the popularity of VR content, which is used to weight tiles. The system integrates content popularity, rate distortion and the information of base stations to generate packet scheduling for resource allocation between multiple base stations. The system, however, cannot realize the viewport of a user in practice since it only uses a frequency model to weight tiles while ignoring users' differences. 
Ge et al. \cite{7997740} propose a multi-path cooperative route scheme for XR transmission. To stream massive VR data with low system delay, the data is repeatedly stored in multiple edge data centers (EDCs). The MCR scheme selects EDCs to meet delay constraints. However, their work aims to determine suitable paths for various VR packets. Want et al. \cite{10494038} propose a viewpoint-aware packet scheduling strategy based on tile-weighted rate-distortion information to reduce data volume and optimize XR streaming under adverse network conditions. The system considers the viewpoint's importance and keeps the high quality of the viewport by the effect of the transmission network.

\begin{figure}[h]
  \centering
  \includegraphics[width=\linewidth]{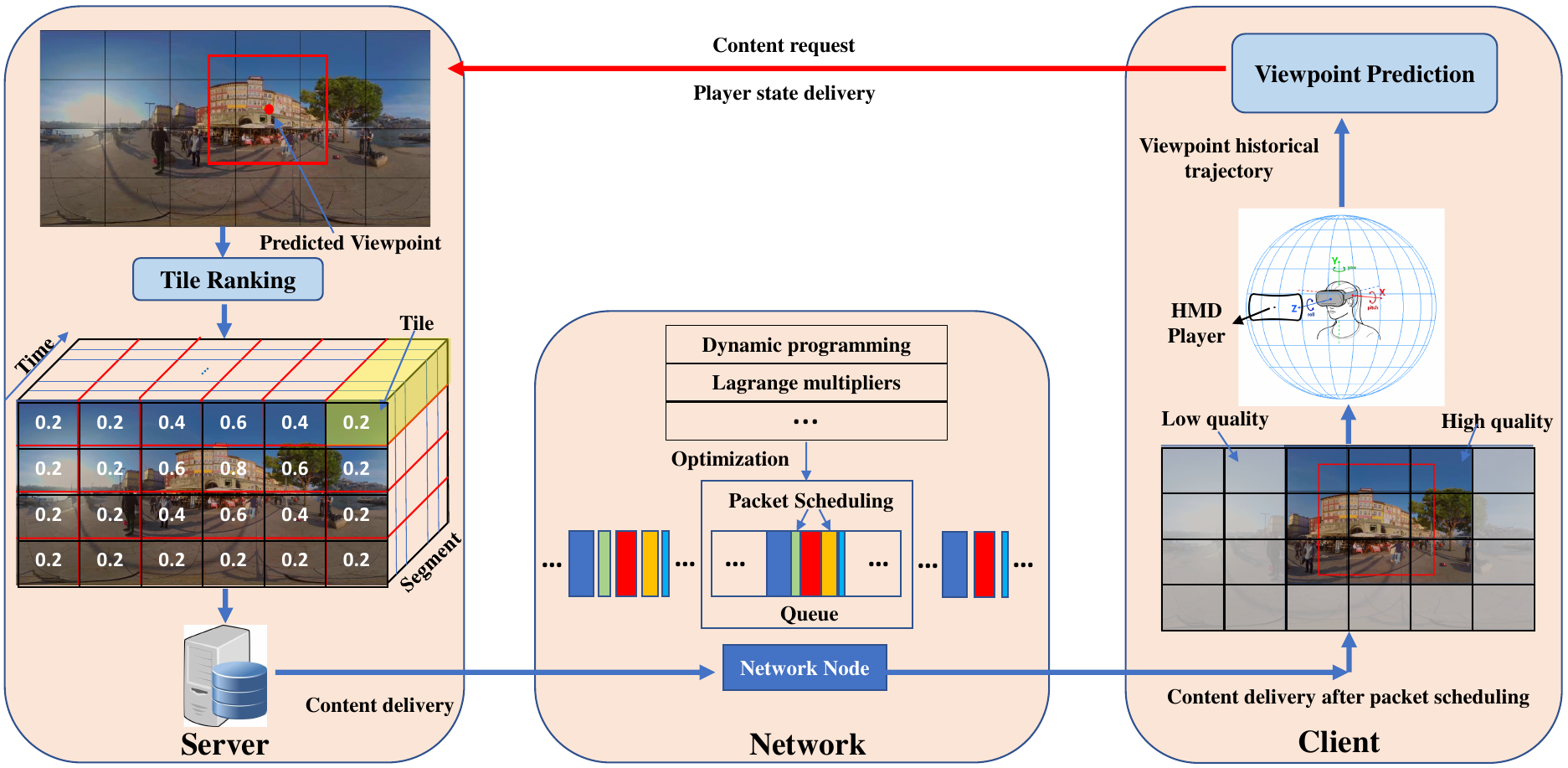 }
  \caption{A typical framework of visual-attention-based packet scheduling method. The streamed packets are scheduled according to the importance of XR content based on viewpoint.}
  \label{packet scheduling}
\end{figure}

\nomenclature{EDC}{Edge Data Center}

The discussed packet scheduling techniques for conventional and XR material are discussed in detail in Table \ref{one}. XR streaming techniques are less researched than traditional content streaming techniques. Current XR packet scheduling methods are more focused on allocating resources among various traffic streams and paths than they are on decreasing the amount of data in a network. In addition, the majority of XR packet scheduling techniques are unable to account for the transmission network's viewport significance. 

Packet scheduling can be formulated as various optimization problems. The rate-distortion optimization problem is a typical one, where the objective is to minimize the transmission rate subject to a constraint on the distortion (or minimize the distortion subject to a transmission rate). As shown in Fig. \ref{packet scheduling}, packets are assigned weights based on the importance of their corresponding XR content. This importance reflects the quality distortion that would result from dropping the packet. The more important a packet is, the higher the distortion it causes. Consequently, the packet scheduling strategy prioritizes maintaining high quality within the viewport by dropping packets outside the viewport when network bandwidth is constrained. The objective function is given by:

\begin{equation}
\min_{\{P_i\}} \sum_{i} P_i D_i
\end{equation}

subject to:
\begin{equation}
\sum_{i} P_i R_i \leq R_{\max}
\end{equation}
\begin{equation}
\sum_{i} P_i = 1
\end{equation}
\begin{equation}
0 \leq P_i \leq 1
\end{equation}
where \( P_i \) is the probability of scheduling the \( i \)-th packet, \( D_i \) is the distortion if the \( i \)-th packet is not transmitted, \( R_i \) is the rate required to transmit the \( i \)-th packet, and \( R_{\max} \) is the maximum allowable transmission rate. To solve this constrained optimization problem, Lagrange multipliers can be used \cite{9086630}.

\section{Ubiquitous XR Applications} \label{6}
The applications of XR span across various industries, leveraging immersive capabilities to enhance experiences, improve efficiencies, and create new opportunities. Research shows that XR can be applied effectively across diverse domains, yielding significant practical benefits. In healthcare, XR is used for surgical training and planning, allowing medical professionals to practice procedures in a risk-free virtual environment. For example, during the COVID-19 pandemic, doctors used XR to provide remote care, enhancing patient outcomes and safety \cite{VRmedical}. 
In the retail sector, companies, such as IKEA and Sephora, have integrated AR into their customer experiences, enabling users to visualize furniture in their homes or experiment with virtual makeup applications, thereby improving customer engagement and satisfaction \cite{IKEA}. In this section, we discuss some key XR applications.
% \begin{figure}[h]
%   \centering
%   \includegraphics[width=\linewidth]{applications.pdf}
%   \caption{The main applications of XR.}
%   \label{applications}
% \end{figure}

\begin{itemize}
    \item \textbf{\textit{Game and Entertainment:}} As XR provides an immersive environment and multimodal interactions, it significantly enhances gaming and entertainment experiences, which allows users to virtual environments and engage with characters and items in a highly realistic environment. Many studies have demonstrated that XR offers a better user experience for gaming and entertainment \cite{gunkel2023immersive, coronado2023integrating, freina2015literature}. VR headsets offer visually and emotionally engaging experiences, allowing users to enter entirely virtual worlds \cite{freina2015literature}. In addition, XR promotes physical health by increasing physical activity. For example, Pokemon Go is an AR game that integrates advanced mobile technology with real-world exploration. Since 2016, it has become a global phenomenon that shocked the world. As demonstrated by Pokemon GO, AR games can increase physical activity and exercise among users by enabling users to interact with digital characters and objects in their real-world surroundings \cite{chong2018going}. Similarly, MR also plays a major role in game and entertainment systems \cite{cheok2009mixed}. The XR will allow humans to interact with each other in ways that have been beyond our imaginations, and the scale of interaction with computers will far surpass what we're used to in today's desktop computers.  
    
    \item \textbf{\textit{Healthcare:}} The XR technologies are revolutionizing various aspects of healthcare, from training and procedure simulation to treatment and rehabilitation, which offers transformative solutions for both patients and healthcare professionals \cite{gahelot2024systematic, lungu2021review}. XR has been adopted in many healthcare fields, such as mental health \cite{pons2022extended, stone2020extended}, physiotherapy \cite{khan2023clinical}, pharmaceutical development \cite{spyrou2023virtual}, and medical education \cite{lungu2021review, logeswaran2021role}. Some healthcare organizations have used XR to train doctors on complex clinical procedures such as simulated surgery. In this way, professionals and students can practice complex procedures in a risk-free environment, thereby gaining confidence and skill \cite{logeswaran2021role}. Furthermore, XR transforms medical imaging by integrating with traditional modalities, such as computed tomography (CT) and magnetic resonance imaging (MRI) scans. This integration provides medical professionals with three-dimensional visualizations of anatomical structures, thus enhancing diagnostic accuracy and facilitating a deeper understanding of complex anatomies \cite{allison2020breast3d}. The integration of XR technologies in healthcare not only showcases its versatility but also its significant potential to transform patient care and improve healthcare outcomes. For example, VR is effectively used for pain management, where patients engage with virtual environments to distract themselves during painful procedures \cite{kaya2023effect, norouzkhani2022effect}. 
    
    \item \textbf{\textit{Education and Training:}} By utilizing the potential of immersive, interactive, and experiential learning settings, XR technologies can provide services and usages that are more solutions-inclusive in education and training. This can aid users in understanding complicated concepts and enhancing their problem-solving abilities. For example, VR offers fully immersive experiences where learners can engage in simulated scenarios, such as historical reenactments or complex scientific experiments, without the constraints of physical reality. Institutions, such as Stanford University, have implemented VR programs that allow medical students to practice surgery in a controlled and safe environment, leading to improved skill acquisition and confidence \cite{plackett2022effectiveness}. Similarly, AR applications such as AR Circuits, enable students to visualize and interact with electronic components and circuits in real-time, enhancing their understanding of electrical engineering principles through interactive, 3D representations \cite{kreienbuhl2020ar}. Furthermore, MR technologies are used in architecture and design courses, where students can manipulate and interact with 3D models of their projects, facilitating a deeper understanding of spatial relationships and design concepts. These XR technologies support active learning and can significantly improve student engagement, motivation, and understanding of complex subjects by offering hands-on experiences that traditional methods may lack \cite{radianti2020systematic}.
    
    \item \textbf{\textit{Travel and Tourism:}} XR can provide realistic virtual tours of tourist destinations, museums, and historical sites, which allows individuals to experience travel destinations without incurring the costs of transportation, accommodation, and other travel expenses. XR technologies also offer potential travelers the opportunity to experience destinations virtually before visiting them physically, aiding in travel planning and decision-making. For instance, XR users can take virtual tours of hotels, attractions, and entire cities, which offers a realistic preview that influences their travel choices \cite{guttentag2010virtual, tussyadiah2018virtual}. Additionally, the adoption of XR in travel and tourism not only improves the customer experience but also offers significant benefits to the industry. By providing virtual experiences, travel businesses can reduce the environmental impact of physical travel, thereby promoting sustainable tourism practices \cite{beck2019virtual}. Moreover, XR technologies serve as powerful marketing tools, helping destinations differentiate themselves in a competitive market with unique, memorable experiences that attract visitors.  For example, the tourism board of the Faroe Islands launched a successful remote tourism campaign using VR, allowing potential tourists to explore the islands through a local guide's perspective, thereby increasing interest and bookings \cite{REMOTETOURISM}. As XR technologies continue to evolve, their integration into tourism is expected to grow, providing new opportunities for innovation and enhancing the overall travel experience.
    
    \item \textbf{\textit{eCommerce and Retail:}} XR technologies are transforming the way consumers engage with products and brands with immersive and interactive experiences. VR creates fully digital environments where users can explore virtual stores or product showrooms, offering a novel way of online shopping. For instance, IKEA's VR showroom allows customers to visualize furniture in a simulated home setting, enhancing the decision-making process \cite{dragan2018state}. Similarly, AR applications, such as Smart Mirror developed by Sephora, enable customers to virtually try on makeup using their smartphones, which provides a personalized shopping experience without requiring a physical store visit \cite{caboni2019augmented}. Additionally, XR technologies are also applied in supply chain management and staff training, providing realistic simulations for training purposes and optimizing warehouse operations \cite{papagiannidis2013modelling}. As XR continues to evolve, its integration into eCommerce and retail is poised to offer increasingly sophisticated and personalized shopping experiences, reshaping the industry's landscape.
    
    \item \textbf{\textit{Engineering and Manufacturing:}} XR is revolutionizing the engineering and manufacturing fields by enhancing visualization, prototyping, and training processes. XR technologies allow engineers to explore and manipulate complex designs in unprecedented ways, which is particularly advantageous in complex engineering and manufacturing environments \cite{ chu2024extended,han2022generic}. XR facilitates the creation of detailed virtual prototypes and enables engineers to conduct comprehensive analyses and modifications in a simulated environment, which significantly reduces the reliance on physical prototypes and shortens the design cycle \cite{nee2012augmented}. For example, in the automotive industry, companies, such as Ford, have leveraged VR to enhance vehicle design processes, allowing engineers to perform virtual walkthroughs of new models and make real-time adjustments based on simulated feedback \cite{ford}. On the other hand, manufacturing and engineering often involve hazardous tasks. XR enables workers to perform these tasks remotely, ensuring their safety \cite{stanney2021extended}. In addition, XR allows teams to collaborate in a shared virtual workspace, improving communication and coordination, and fostering innovative problem-solving \cite{pantelidis2009reasons}. These advancements not only streamline workflows but also enhance the overall productivity and safety of manufacturing operations. 
\end{itemize}

\section{Challenges and Future Research Directions} \label{7}
Extensive research has been devoted to enhancing XR performance in networks with limited bandwidth and high variability. However, streaming high-quality XR content remains a significant challenge. This paper highlights several promising research challenges and explores potential research directions for advancing this field.

\begin{itemize}
    \item \textbf{\textit{Real-time Rendering and Transmission:}} Real-time rendering and transmission in XR streaming present substantial challenges, particularly in balancing the requirements for low latency, high bandwidth, and computational efficiency. Low latency is crucial for maintaining immersive experiences, as even minor delays can disrupt the user's sense of presence and lead to discomfort. Research on the Tactile Internet, which targets latencies below 10 milliseconds, along with advancements in network technologies such as 5G, are critical in addressing these latency challenges. Thus, effectively managing network congestion and adapting to network heterogeneity, including variations in Wi-Fi and 5G connectivity, are imperative for ensuring consistent performance. Additionally, minimizing jitter and packet loss is essential for preserving a seamless user experience. Simultaneously, delivering high-resolution 3D content, especially in multi-user environments, demands significant bandwidth. This necessitates the development of advanced compression algorithms and adaptive streaming techniques to manage bandwidth efficiently without compromising the QoE. Moreover, the real-time rendering of complex scenes requires considerable computational power, placing heavy demands on GPUs. High-performance GPUs, such as those developed by NVIDIA, are essential to meeting these needs, but further optimization remains necessary. This includes improving hardware architectures and developing more efficient algorithms to ensure that XR devices, particularly portable and wearable ones, can handle these computational demands while maintaining energy efficiency.
    
    \item \textbf{\textit{QoE:}} The QoE in XR streaming is a multifaceted concept encompassing both technical and experiential elements to deliver immersive user experiences. Effective QoE models must address the unique challenges of XR, such as ensuring visual fidelity even with necessary data compression, enhancing immersion through sophisticated spatial audio processing, and integrating emerging technologies, such as haptic, smell and taste feedback, to engage additional senses for a more holistic experience. Natural interaction methods are essential, necessitating intuitive yet unobtrusive designs that enable seamless user interaction with XR environments. User comfort is another essential aspect, with particular focus on mitigating motion sickness and fatigue to support extended use of XR devices. Social interaction within virtual spaces introduces additional complexity, demanding sophisticated collaboration tools that address both technical and social dynamics. While the field of QoE in XR is still developing, a thorough understanding of these interconnected factors is vital for creating robust models that significantly enhance user experiences in XR environments.

    \item \textbf{\textit{Viewpoint Prediction:}} A fundamental challenge in XR streaming is viewpoint prediction. Despite extensive research on this topic, existing approaches often yield inaccurate viewpoint prediction results. Users' attention in XR environments is shaped by various factors, including individual behavior, past movements, video content, and external influences. Consequently, deep learning-based solutions will be essential for future advancements. The ability to accurately forecast long-term attention patterns in various complex scenarios is beyond the capability of even the most advanced models. Currently, most learning-based methods have the capability to accurately forecast viewpoint trajectory for a duration of up to 5 seconds. It is crucial to increase the duration of this forecast period in order to enable other elements of the encoding and streaming process to adapt smoothly to the user's real-time behavior.
    
    \item \textbf{\textit{Standards and Protocols:}} As an emerging field, XR lacks universally accepted standards and protocols, leading to fragmentation in hardware, software, and content. This fragmentation creates compatibility issues across different platforms and devices, hindering seamless user experiences and content interoperability. Furthermore, the lack of standardized development frameworks and communication protocols complicates the integration of XR applications with existing technologies and systems. Ensuring consistent quality, performance, and security across diverse XR environments further complicates development efforts. Moreover, the unique traffic and performance characteristics of XR demand significant improvements in standards and protocols. XR’s hardware and content requirements, including high resolutions, frame rates, interactivity, mobility, and novel data types, such as haptics and spatial tracking, pose additional challenges. To address these challenges, industry and academic stakeholders must work together to develop and implement thorough standards and protocols that will enable the expansion and scalability of XR technology. 
    
    \item \textbf{\textit{Lightweight XR Solutions:}} Current XR solutions rely on wearable devices, which are often costly and inconvenient. XR HMDs incorporate a built-in processor and battery, making them cumbersome and heavy to wear. Consequently, existing HMDs cannot offer a lightweight yet high-quality XR experience. There is an urgent need for the design of a lightweight XR solution that can deliver an optimal user experience without the bulk and inconvenience of current devices.
    
    \item \textbf{\textit{Environmental Mapping:}} Environmental mapping is another area fraught with challenges. XR applications must accurately map and interpret the user’s environment to provide relevant and immersive experiences, which necessitates advanced algorithms for spatial recognition and tracking. Rendering objects accurately under different lighting conditions and ensuring correct occlusion by displaying objects in the proper spatial order, either in front of or behind others, are complex technical challenges.
    
    \item \textbf{\textit{Content Creation and Management:}} Content creation and management represent a critical challenge in XR streaming, necessitating a careful equilibrium between the production of high-quality, realistic graphics and the maintenance of optimal performance. The intensive demands of rendering detailed graphical assets can impose significant strain on hardware resources. Furthermore, the creation of high-fidelity 3D models, animations, and immersive environments is inherently time-consuming, requiring specialized expertise, often resulting in developmental bottlenecks. To mitigate these challenges, the deployment of advanced authoring tools and workflows is indispensable. The integration of Large Language Models (LLMs) offers a promising avenue for enhancing efficiency in generating complex, interactive 3D worlds and haptic environments, thereby alleviating the workload on content creators. Nevertheless, the distribution of this content across devices with varying capabilities remains a formidable obstacle, necessitating adaptive strategies to ensure a consistent and high-quality user experience. Additionally, the effective management and organization of extensive immersive content are imperative for ensuring discoverability. In this context, metadata management, augmented by LLMs, plays a vital role in enhancing the searchability of devices and content, thereby facilitating users' access to desired immersive experiences.

\end{itemize}
Given the substantial challenges XR technology faces, particularly around bandwidth limitations, computational demands, and latency, several innovative solutions are under exploration. Real-time rendering and high-quality streaming require substantial bandwidth, posing significant challenges to existing network infrastructures. To address these issues, adaptive streaming techniques, such as foveated rendering, allocate high-resolution processing on the user's gaze area, thereby reducing overall computational load. Additionally, the deployment of 5G networks offers lower latency and higher data transfer rates, facilitating more seamless XR experiences. Edge computing and cloud-based rendering are other viable solutions, processing data closer to the XR devices to reduce latency and offload processing demands from the devices themselves. These approaches collectively address the technical challenges of XR, facilitating its broader adoption across various sectors.

\section{conclusion} \label{8}
Due to improvements in network bandwidth and computational capacity, people are now demanding more immersive XR experiences. This survey presents a comprehensive analysis of the latest research on XR streaming, aiming to bridge gaps left by prior studies that focused narrowly on 360-degree video or specific domains, such as education and healthcare, by exploring a broad range of topics, from XR systems to applications that have yet to be thoroughly examined. Since XR content is distinct from traditional media, analyzing XR traffic characteristics is crucial to understanding its unique network infrastructure requirements. We provide a detailed analysis of traffic patterns, device architectures, multimodal interactions, and adaptive streaming technology. Additionally, we analyze the factors influencing XR QoE to ensure systems meet user expectations and deliver compelling, immersive experiences.
This survey also explores advanced optimization strategies at both the application and network layers, addressing challenges and future research directions. By highlighting the unique challenges and QoE demands of XR streaming, the paper provides foundational insights for future developments in immersive multimedia communication. Ultimately, our goal is to inspire innovative research in XR streaming and ultimately improve immersive XR experiences in everyday life.

\bibliographystyle{ACM-Reference-Format}
\bibliography{sample-base}

\appendix
\section{Appendix}
\printnomenclature

\end{document}